\documentclass[fleqn,usenatbib]{mnras}

\usepackage{newtxtext,newtxmath}

\usepackage[T1]{fontenc}

\DeclareRobustCommand{\VAN}[3]{#2}
\let\VANthebibliography\thebibliography
\def\thebibliography{\DeclareRobustCommand{\VAN}[3]{##3}\VANthebibliography}

\usepackage{graphicx}	
\usepackage{amsmath}	
\usepackage{longtable}
\usepackage{lscape}

\title[Extragalactic transient candidates in the 2SXPS catalogue]{Extragalactic transient candidates in the Second \textit{Swift}-XRT Point Source catalogue}

\author[R. A. J. Eyles-Ferris et al.]{
R. A. J. Eyles-Ferris,$^{1}$\thanks{E-mail: raje1@leicester.ac.uk}
R. L. C. Starling,$^{1}$
P. T. O'Brien$^{1}$
and P. A. Evans$^{1}$
\\
$^{1}$School of Physics and Astronomy, University of Leicester, University Road, Leicester, LE1 7RH, UK\\
}

\date{Accepted XXX. Received YYY; in original form ZZZ}

\pubyear{2022}

\begin{document}
\label{firstpage}
\pagerange{\pageref{firstpage}--\pageref{lastpage}}
\maketitle

\begin{abstract}
The Second \textit{Swift}-XRT Point Source catalogue offers a combination of sky coverage and sensitivity and presents an invaluable opportunity for transient discovery. We search the catalogue at the positions of inactive and active galaxies, and identify transient candidates by comparison with \textit{XMM-Newton} and \textit{ROSAT}. We recover 167 previously known transients and find 19 sources consistent with being new sources, estimating a completeness of $\sim65\%$. These 19 new sources are split approximately equally between inactive and active hosts and their peak X-ray luminosities span $\sim 10^{42} - 10^{47}$ erg s$^{-1}$. We find eight are best fit with non-thermal spectral models and one with a blackbody. We also discuss our methodology and its application to the forthcoming Living \textit{Swift}-XRT Point Source catalogue for the potential near real time serendipitous discovery of $\sim$ few new X-ray transients per year.
\end{abstract}

\begin{keywords}
catalogues -- X-rays:general -- transients
\end{keywords}



\section{Introduction}

The Second \textit{Swift}-XRT Point Source\footnote{\url{https://www.swift.ac.uk/2SXPS/}} \citep[2SXPS,][]{Evans19} catalogue has been constructed from hundreds of observations undertaken by the  \textit{Neil Gehrels Swift Observatory}, hereafter \textit{Swift}, using its X-ray Telescope (XRT). The timing of these observations ranges from early in \textit{Swift}'s mission, 1 January 2005, up until 1 August 2018.

2SXPS has been constructed both from individual images and from stacks of images to maximise the S/N ratio. In the XRT's full 0.3--10 keV band, the 2SXPS achieves a sensitivity of $1.73\times10^{-13}$ erg cm$^{-2}$ s$^{-1}$ for individual pointings and $\sim4\times10^{-14}$ erg cm$^{-2}$ s$^{-1}$ for stacked images with a sky coverage of 1790 deg$^2$. Such a depth is $\sim1-3$ orders of magnitude deeper than the  \textit{ROSAT} all sky and \textit{XMM-Newton} slew surveys, while the sky coverage is a $>3$ times larger area than 4XMM-DR11. Such a combination of depth and coverage is currently unique even compared to 2SXPS' predecessor 1SXPS and 2SXPS therefore offers an excellent opportunity to explore the high energy sky.

While \textit{Swift} has proven to be excellent at both discovery and follow-up of transient events, 2SXPS represents a chance to further expand on those results. The large field of view of the XRT means that a number of additional transients could be detected in the background of observations they were not the focus of. Such sources wouldn't necessarily be immediately identified as transients but the construction methodology of 2SXPS ensures they will be included in this catalogue. Interrogating 2SXPS could therefore lead to identification of these transient sources.

Such work has also been undertaken for other X-ray catalogues. For example, the \textit{XMM-Newton} Slew Survey catalogues \citep[XMMSL1 and XMMSL2, ][]{Saxton08,Warwick12} were examined in this way. By comparing the Slew Survey fluxes to the \textit{ROSAT} all sky survey counterparts or upper limits, a significant number of transient candidates were identified including several tidal disruption events \citep[TDEs,][]{Saxton12,Saxton17,Saxton17a,Li20,Li22}. The presence of further transient candidates in the Slew Survey was also inferred by later non-detections using \textit{Swift} \citep{Starling11}.

In this paper, we present our search for such transient candidates in the 2SXPS catalogue. In Section \ref{sec:samples}, we detail our procedure for selecting 2SXPS sources coincident with galaxies to target extragalactic transients, then discuss how flaring candidates were identified in these samples in Section \ref{sec:flare_id}. The candidate sources are further examined and initially classified in Section \ref{sec:src_class} and the most likely transient candidates are characterised in Section \ref{sec:transient_character}. Finally, in Section \ref{sec:discussion}, we discuss our methods, particularly in regards to the future live version of the \textit{Swift}-XRT Point Source catalogue, and present our conclusions in Section \ref{sec:conc}.

Throughout this paper we adopt a cosmology with $H_0 = 71$ km\,s$^{-1}$\,Mpc$^{-1}$, $\Omega_m = 0.27$ and $\Omega_\Lambda = 0.73$.

\section{Source samples}
\label{sec:samples}

\begin{figure*}
	\includegraphics[width=\textwidth]{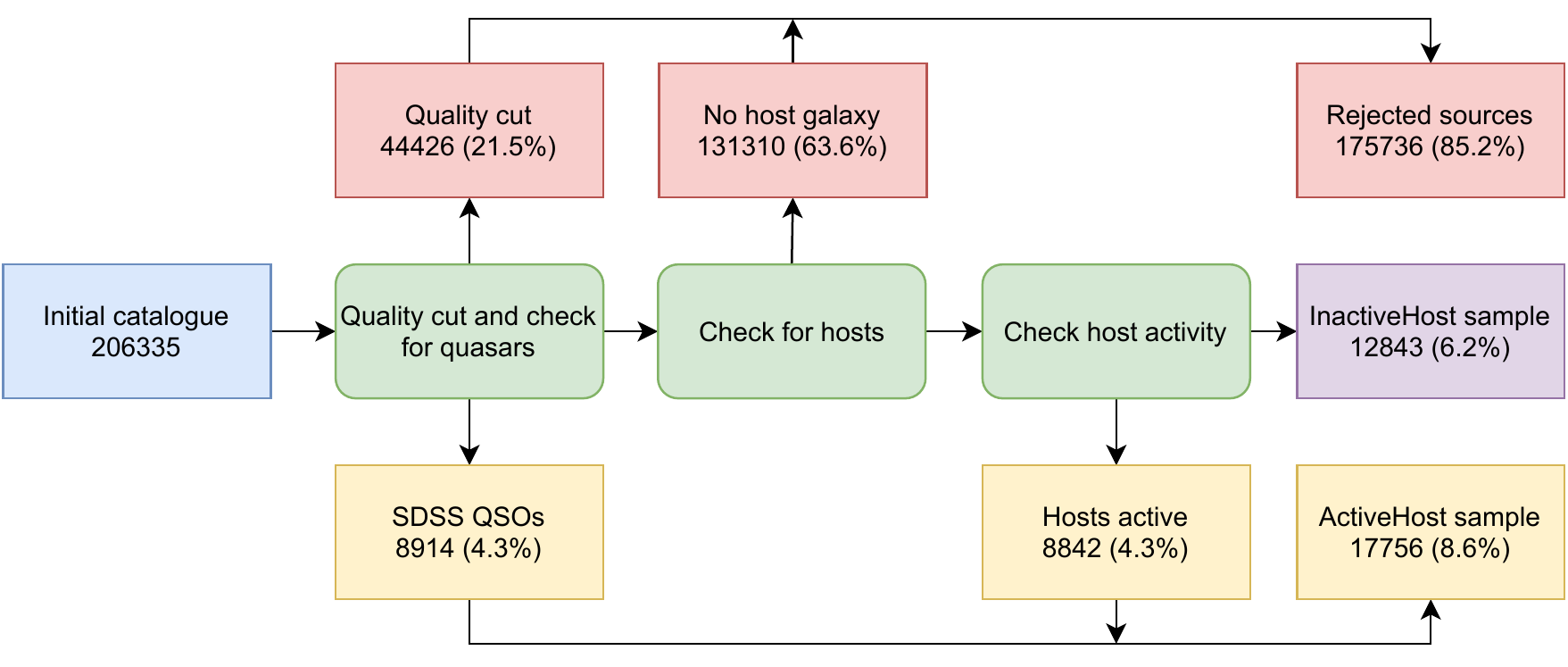}
    \caption{A flowchart illustrating the catalogue cuts and cross-correlations used to identify and define the InactiveHost and ActiveHost samples. Blue indicates the initial catalogue, green indicates the processes applied, red indicates rejected sources, yellow indicates ActiveHost sources and purple indicates InactiveHost sources.}
    \label{fig:flowchart}
\end{figure*}

The 2SXPS catalogue consists of some 206,335 sources. To efficiently identify transient candidates, we therefore had to devise suitable smaller samples from the overall catalogue. To maximise the chance of any transients being extragalactic, we selected sources that were colocated with known galaxies. Two such samples were produced, consisting of sources apparently hosted in inactive galaxies (the InactiveHost sample) and those hosted in galaxies where an active galactic nucleus (AGN) was also present (the ActiveHost sample). It can be common, for example in the case of TDE searches, to exclude sources found to be hosted in active galaxies \citep{vanVelzen11}. However, the X-ray luminosity of AGN is expected to vary only by a factor of a few \citep[e.g.][]{Maughan19} while transient events are likely to cause significantly greater variation. Both samples were therefore included and examined but the ActiveHost sample was subject to stricter scrutiny to minimise AGN contamination. There were several steps in the procedure to produce both samples and an overview of our full process is shown in Figure \ref{fig:flowchart}.

Our first step was to perform a quality cut using 2SXPS' \texttt{DetFlag} parameter, selecting only \textit{Good} or \textit{Reasonable} sources. This reduced the initial sample to 161,909 (78.5\% of the initial catalogue) sources. The 2SXPS pipeline also cross matches to several other catalogues as described in \citet{Evans19}, and the remaining 8,914 (4.3\%) sources that had been successfully matched to the SDSS Quasar Catalogue DR14 \citep{Paris18} were added to the ActiveHost sample. Both checks are summarised by the left hand green box in Figure \ref{fig:flowchart}, and the rejected sources and SDSS quasars are designated in red and yellow respectively.

To identify those sources of the remaining 152,995 (74.2\%) that were most likely to be extragalactic, we cross-correlated them to the NASA/IPAC Extragalactic Database\footnote{\url{https://ned.ipac.caltech.edu/}} (NED), as shown in the middle green box of Figure \ref{fig:flowchart}. We limited our search of NED to galaxies and assumed a source to be hosted in a galaxy if the combined 3-$\sigma$ position error was less than the source-galaxy separation. While this was a fairly generous criterion that would likely lead to spurious matches, it reduced the probability of a real event being rejected. Of the remaining sources, 21,685 (10.5\% of the full catalogue or 14.2\% of 152,995) were found to have putative hosts. Again, rejected sources are designated in red in Figure \ref{fig:flowchart}.

The activity of these hosts was then checked, first by their categorisation in NED. To reduce the impact of incompleteness in NED's categorisations, we also cross matched the host galaxies to both the Million Quasars catalogue\footnote{\url{http://quasars.org/milliquas.htm}} \citep[Milliquas Version 6.4c,][]{Flesch19} and the Quasars and Active Galactic Nuclei catalogue\footnote{\url{https://vizier.u-strasbg.fr/viz-bin/VizieR-3?-source=VII/258/vv10}} \citep[13th Ed.,][]{Veron-Cetty10}. From these checks, shown in the right hand green box in Figure \ref{fig:flowchart}, 8,842 (4.3\%) hosts were found to most likely be active and were therefore added to the ActiveHost sample (yellow in Figure \ref{fig:flowchart}), which consisted of a final 17,756 (8.6\%) sources. The remaining 12,843 (6.2\%) sources made up the InactiveHost sample shown in purple.

Overall, 30,599 (14.8\%) sources were including in the final samples. This relatively small proportion is to be expected as the majority of 2SXPS sources are likely to be stellar sources within the Milky Way.

\section{Flare identification}
\label{sec:flare_id}

From the InactiveHost and ActiveHost samples, we next identified sources that appeared to be flaring. This was done by both analysis of the 2SXPS light curve for each source, and by comparison with external catalogue data.

\subsection{2SXPS light curves}
\label{subsec:lc}

The XRT light curves were acquired from 
the 2SXPS webpages provided by the UK \textit{Swift} Science Data Centre\footnote{\url{http://www.swift.ac.uk/index.php}} (UKSSDC).

The observed count rates were then used to derive the unabsorbed fluxes using \texttt{PIMMS\footnote{\url{https://heasarc.gsfc.nasa.gov/docs/software/tools/pimms.html}} v4.11}. We assumed each source to have the same spectrum, a power law with $\Gamma=1.7$, a typical AGN spectrum. The Galactic absorption column densities were taken from \citet{Willingale13} and the host column densities assumed to be $1\times10^{21}$ cm$^{-2}$ comparable with the mean Milky Way column density. For those sources whose hosts didn't have a catalogued redshift, we assumed the redshifts to be the means of their particular sample, $z=0.326$ and $z=1.082$ for the InactiveHost and ActiveHost samples respectively. These fluxes were used to derive the ratio of the peak flux to the mean flux, $R_{\rm Flux}$, for each source.

The hardness ratios of the sources were also available in addition to the flux data. As many X-ray transients, such as TDEs, exhibit soft spectra, a softening hardness ratio could be indicative of a transient, particularly close to the peak of a light curve. We selected Hardness Ratio 1 (HR1, (Medium - Soft) / (Medium + Soft) where Soft is 0.3--1 keV and Medium is 1--2 keV) as it would best describe the soft region transients would be expected to occupy. The mean value of HR1 was subtracted from the value near the peak to give $\Delta{\rm HR}$. A source softening towards its peak would therefore have a negative value of $\Delta{\rm HR}$. However, in many cases, the sources were not detected enough times for reliable estimates of $\Delta{\rm HR}$ to be derived. For these sources, which consisted of 10349 (80.6\%) InactiveHost sources and 13797 (77.7\%) ActiveHost sources, $\Delta{\rm HR}$ was set to zero.

\begin{figure}
	\includegraphics[width=\columnwidth]{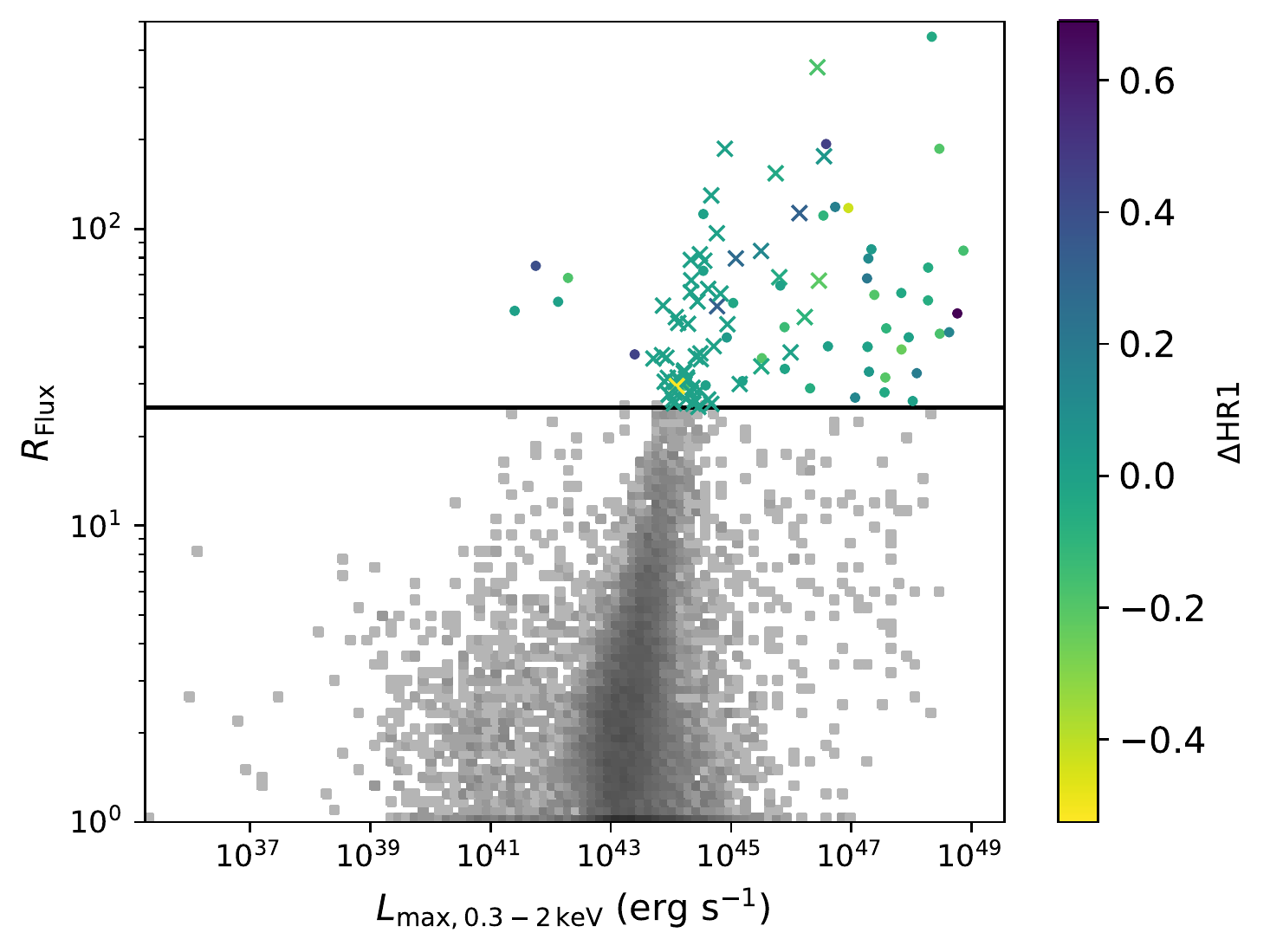}
	\includegraphics[width=\columnwidth]{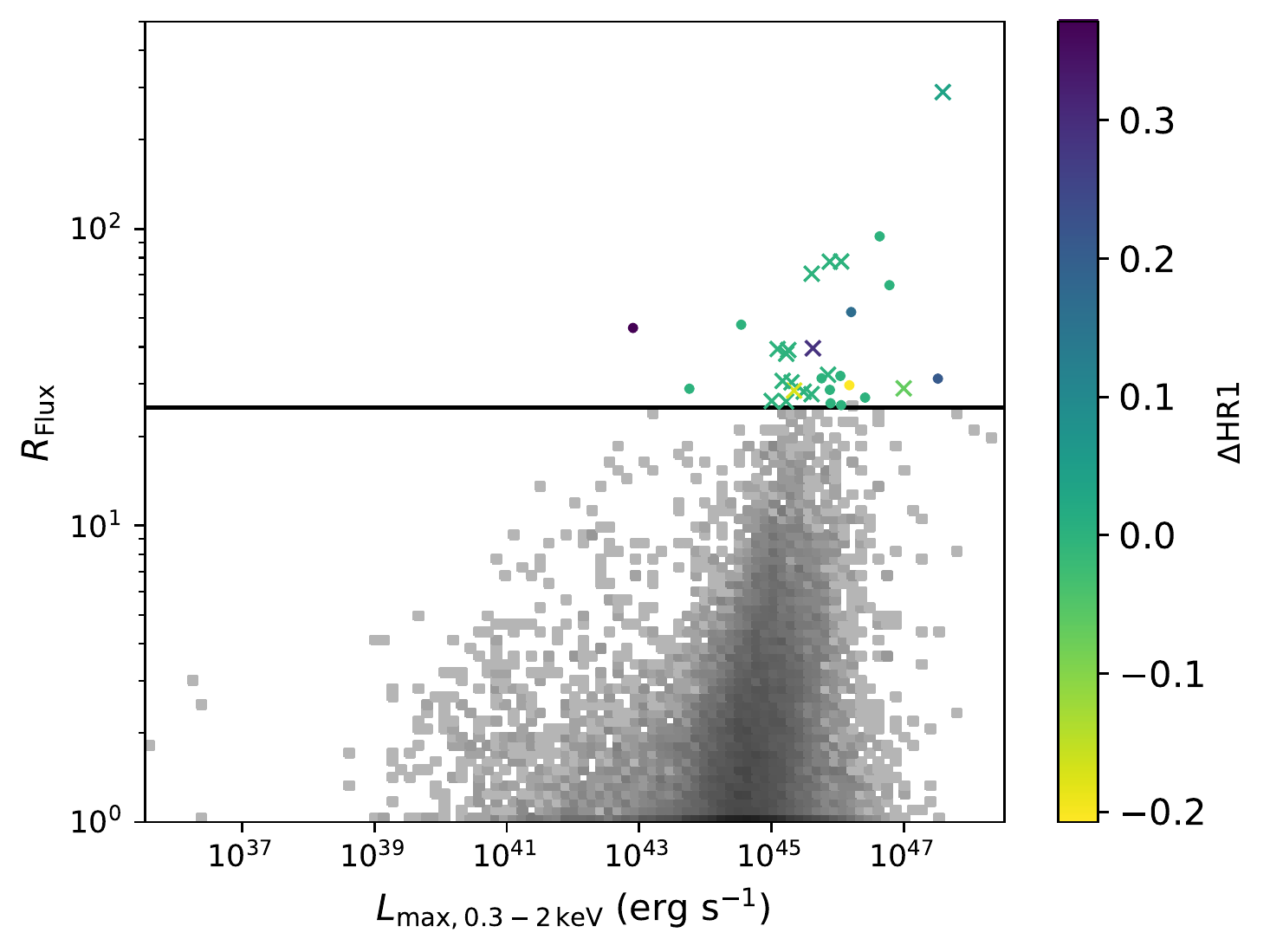}
    \caption{The ratio of peak flux to mean flux in the 2SXPS light curves against X-ray luminosity for the InactiveHost, top, and ActiveHost, bottom, samples. The greyscale is the number density of the majority of the population and coloured sources obey $R_{\rm Flux}\geq25$, also marked by the black horizontal line, while the colour indicates change in hardness ratio. The hosts of sources indicated with circles have measured redshifts whereas those indicated with $\times$ symbols do not and redshifts have been assumed as described in the text. Note also the differing scales on the axes and colour bar.}
    \label{fig:swiftcomp}
\end{figure}

To select flaring sources, $R_{\rm Flux}\geq25$ threshold was set. This threshold was chosen as it selected sources whose behaviour was significantly discrepant from the majority of the population. This threshold indicated 110 (0.9\%) InactiveHost and 31 (0.2\%) ActiveHost sources were flaring. Figure \ref{fig:swiftcomp} shows $R_{\rm Flux}$ against the calculated luminosity of each source, illustrating the divergent behaviour of the selected sources. Note that the calculated luminosities are significantly higher than would be expected from a transient population, which is likely indicative of incorrectly assumed redshifts. However, the luminosity is not used for flare selection so these sources were left in the sample to reduce the risk of rejecting real flares.

\subsection{External catalogue data}

To identify additional flares, the sources were cross-matched to several external catalogues and the peak fluxes compared to the catalogued values. Three catalogues were selected, the 2RXS\footnote{\url{http://vizier.u-strasbg.fr/viz-bin/VizieR?-source=J/A+A/588/A103}} catalogue \citep{Boller16} derived from \textit{ROSAT}'s all sky survey data and two \textit{XMM-Newton} catalogues, the Second \textit{XMM} Slew Survey catalogue\footnote{\url{https://www.cosmos.esa.int/web/xmm-newton/xsa}} \citep{Saxton08} and the 4XMM-DR9\footnote{\url{http://xmmssc.irap.omp.eu/Catalogue/4XMM-DR9/4XMM_DR9.html}} catalogue \citep{Webb20}. 

\subsubsection{2RXS}

2RXS uses the \textit{ROSAT} all sky survey date for the period June 1990 to August 1991. These data were reprocessed to maximise the information gained and a depth of $10^{-13}$ erg cm$^{-2}$ s$^{-1}$ in the 0.1--2.4 keV band was achieved while $\sim135,000$ sources were identified.

In the 2SXPS pipeline cross match to 2RXS, there were found to be 124 (1.0\%) InactiveHost matches and 439 (2.5\%) ActiveHost matches. To estimate \textit{ROSAT }upper limits for the remaining \textit{Swift} sources, we used the Upper Limit Server\footnote{\url{http://xmmuls.esac.esa.int/upperlimitserver/}} to calculate upper limits for a random sample of 100 unmatched sources from 2SXPS. This distribution was found to be relatively narrow and the mean of 0.0313 counts s$^{-1}$ was assumed to be suitably representative of the catalogue's upper limit. The 2RXS 0.1--2.4 keV band count rates were used to derive 0.3--10 keV fluxes using the same spectral shape detailed in \ref{subsec:lc}. The peak fluxes in the 2SXPS light curves were then divided by these values to derive a new measure of $R_{\rm Flux}$.

\begin{figure}
	\includegraphics[width=\columnwidth]{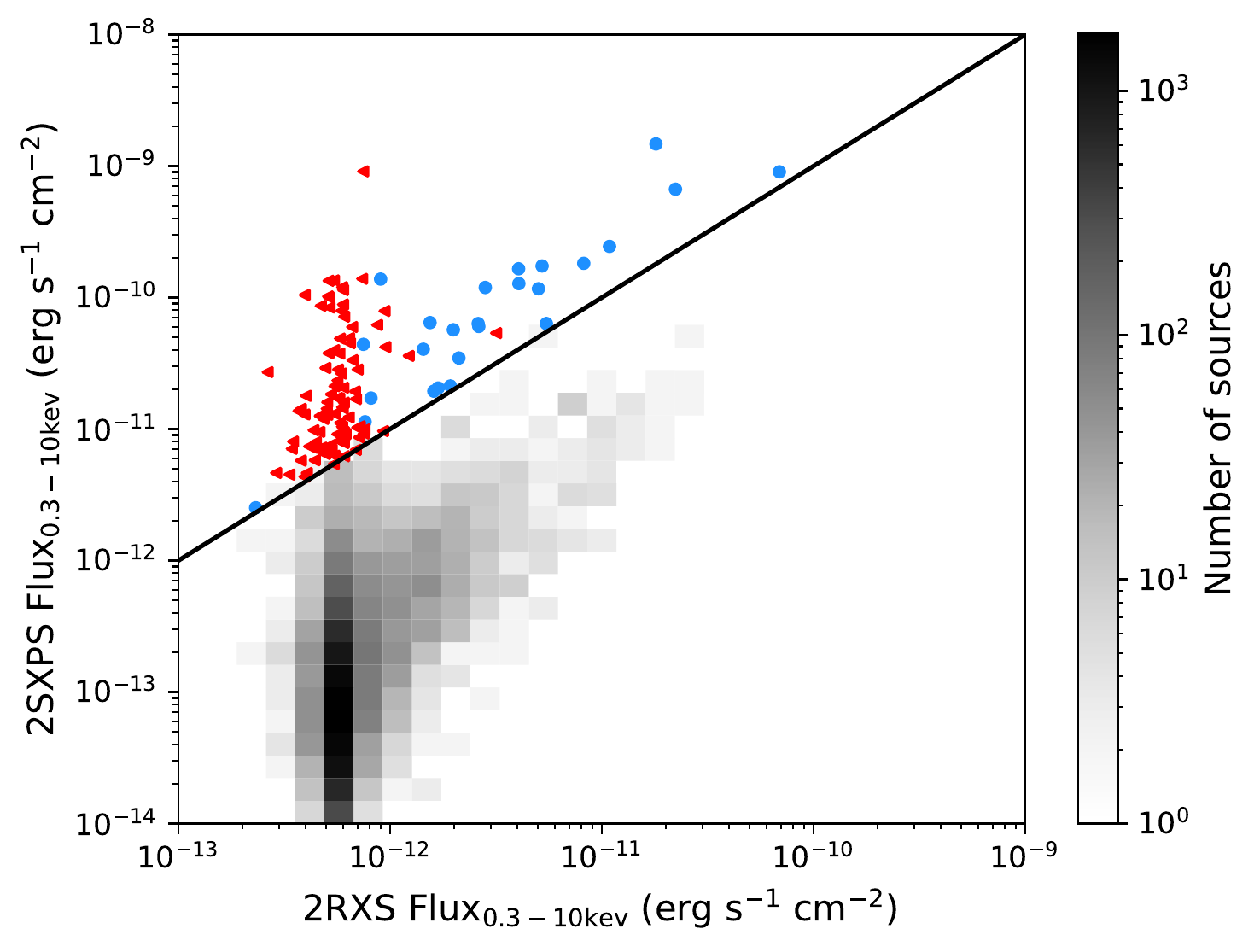}
	\includegraphics[width=\columnwidth]{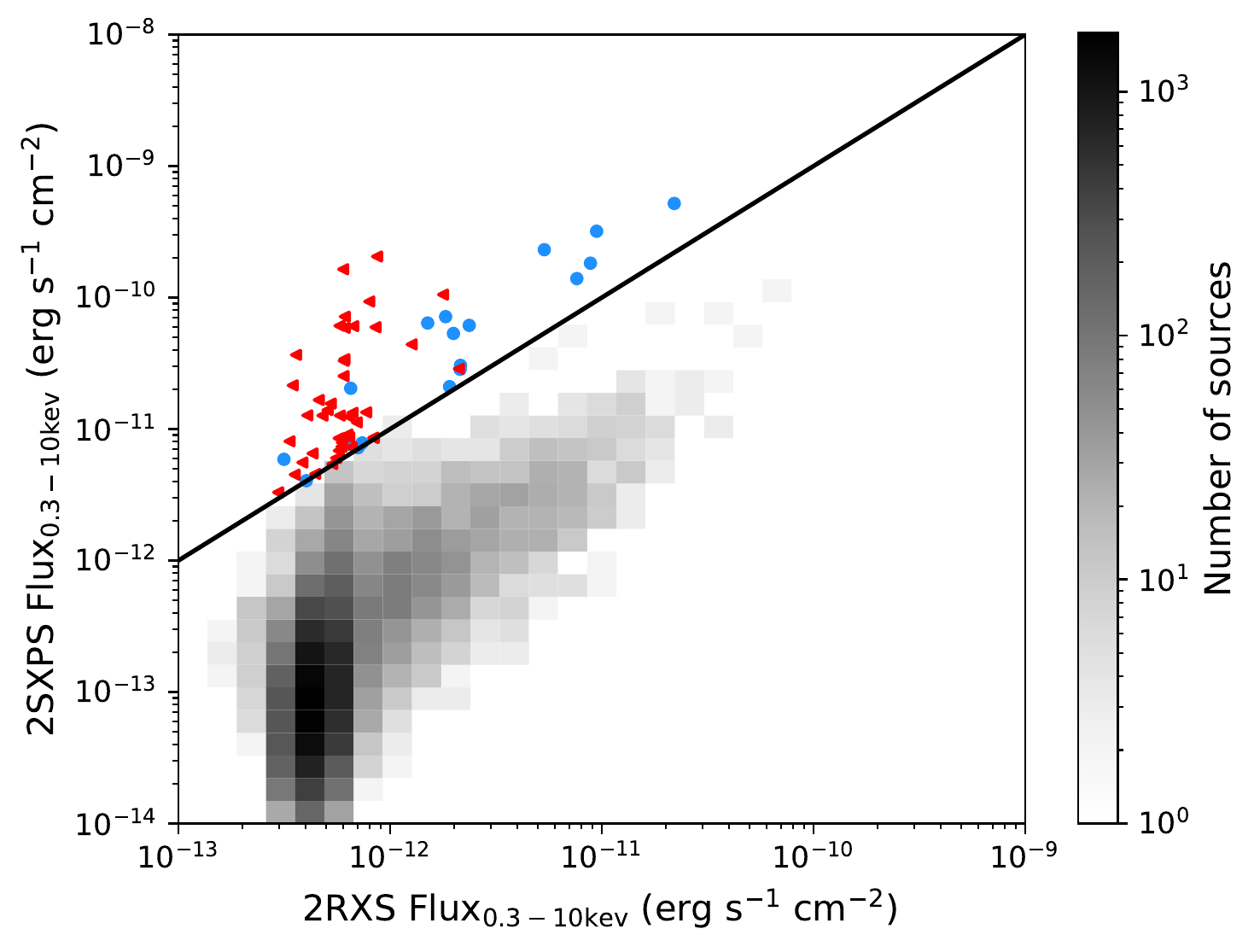}
    \caption{The flux of InactiveHost sources, top, and ActiveHost sources, bottom, in the 2SXPS and 2RXS catalogues. The black line indicates the $R_{\rm Flux}\geq10$ flaring threshold, the greyscale is the number density of the majority of the source population while sources matching the selection criterion are shown with the blue points indicating sources detected in both catalogues and the red triangles indicating sources with only upper limits for \textit{ROSAT}.}
    \label{fig:2rxscomp}
\end{figure}

To account for the effects of Eddington bias due to 2RXS' poorer sensitivity relative to 2SXPS, a lower threshold of $R_{\rm Flux}\geq10$ was chosen, as shown in the black line in Figure \ref{fig:2rxscomp}. This figure also shows that the chosen threshold is effective at selecting only sources with significantly divergent behaviour. This comparison indicated 117 InactiveHost and 63 ActiveHost sources to be flaring at the time of their 2SXPS observations.

\subsubsection{XMMSL2}

While \textit{XMM-Newton} has not performed an all sky survey, it has still produced a wealth of X-ray data. XMMSL2 is constructed from the data observed during slews of \textit{XMM-Newton} and provides the greatest sky coverage of all \textit{XMM-Newton} catalogues. It covers the period 26 August 2001 to 31 December 2014, comparable to 2SXPS, and reaches a sensitivity of $\sim1.2\times 10^{-12}$ erg cm$^{-2}$ s$^{-1}$ in the 0.2--12 keV band and $\sim3\times 10^{-12}$ erg cm$^{-2}$ s$^{-1}$ in the 0.2--2 keV band \citep{Warwick12}.

\begin{figure}
	\includegraphics[width=\columnwidth]{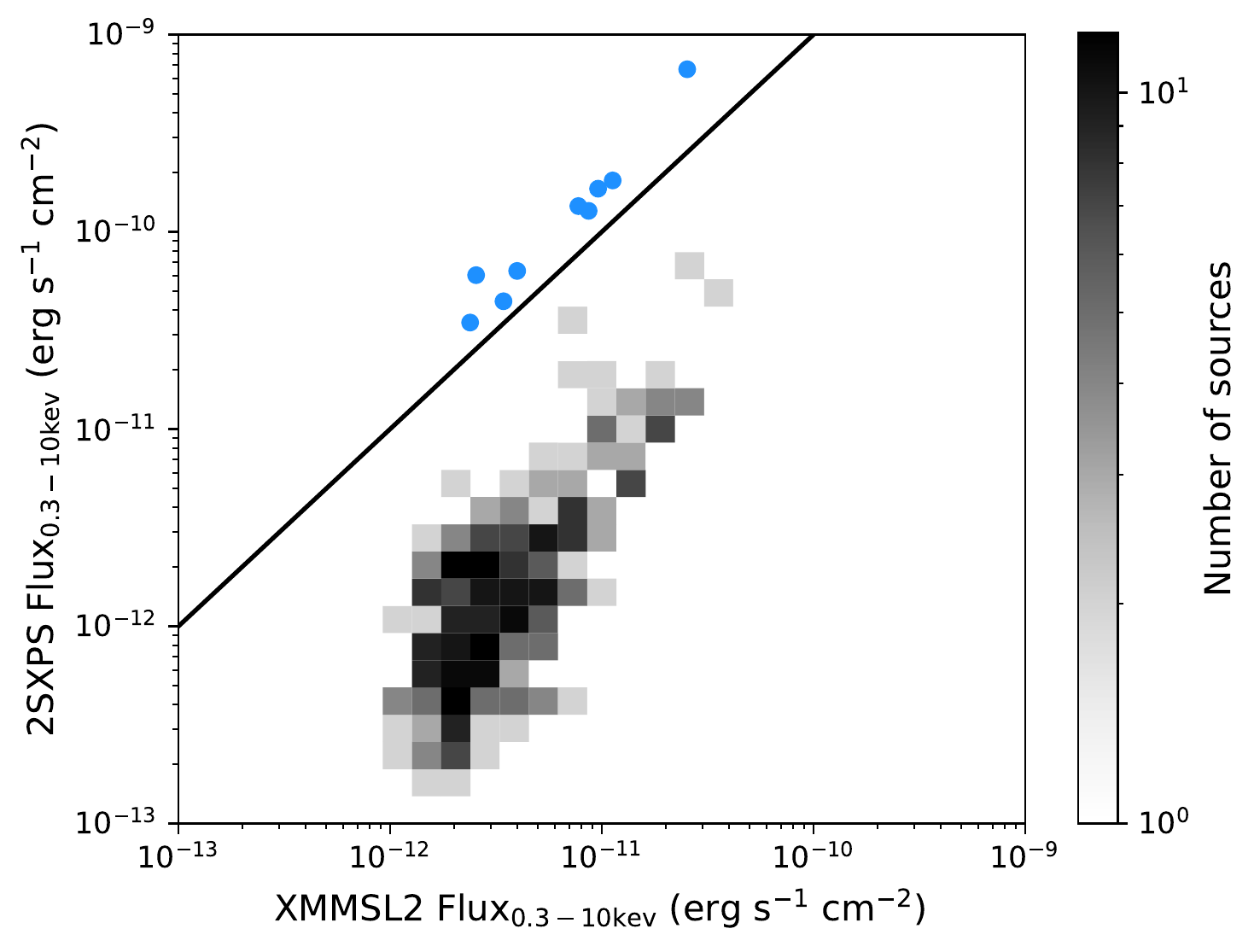}
	\includegraphics[width=\columnwidth]{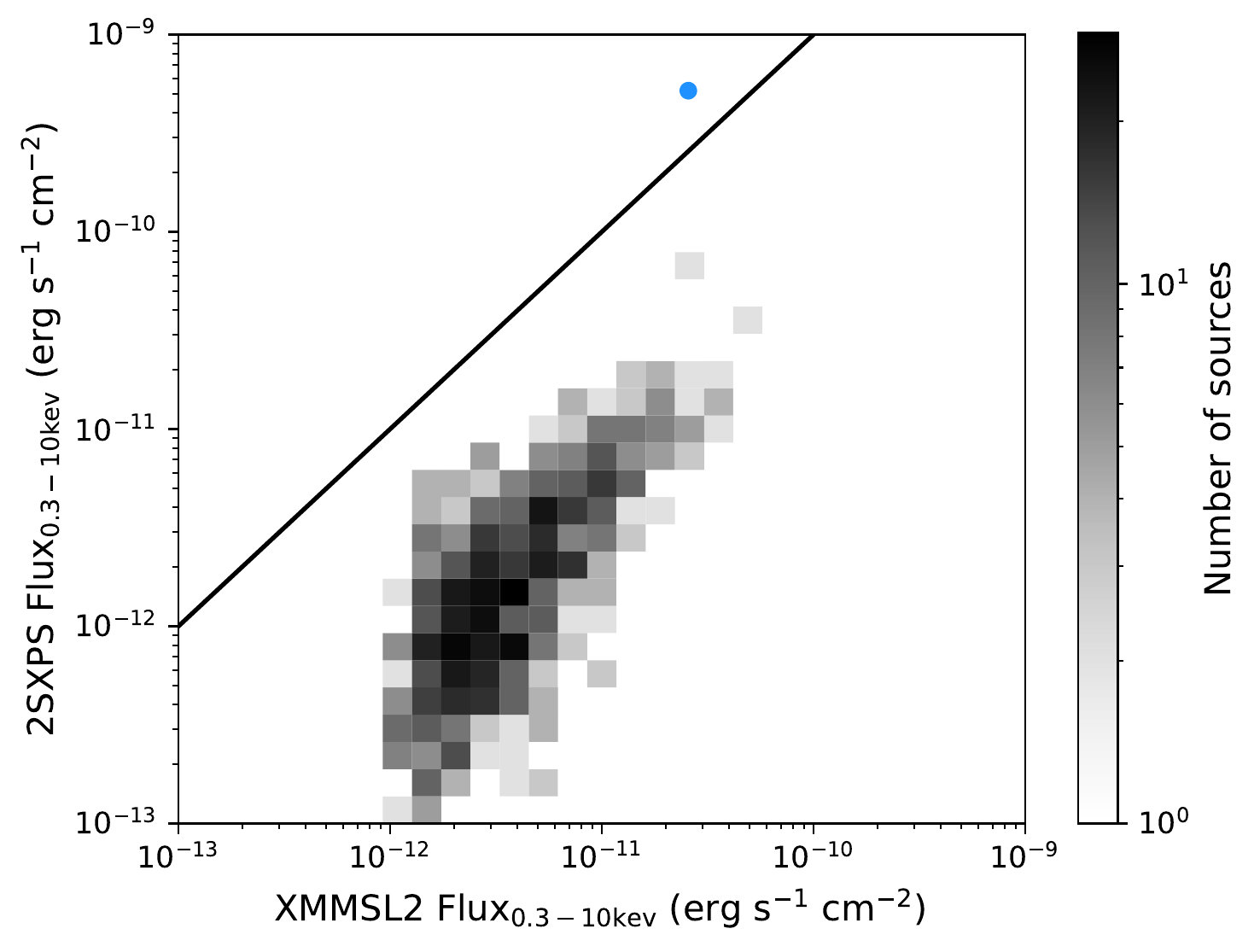}
    \caption{The flux of InactiveHost sources, top, and ActiveHost sources, bottom, in the 2SXPS and XMMSL2 catalogues. The black line indicates the $R_{\rm Flux}\geq10$ flaring threshold, the greyscale is the number density of the majority of the source population and the blue points indicate sources meeting the selection criterion.}
    \label{fig:xmmsl2comp}
\end{figure}

Similarly to 2RXS, the 2SXPS pipeline matches to XMMSL2. This yielded 485 (3.8\%) InactiveHost matches and 1062 (6.0\%) ActiveHost matches. We converted the 0.2--12 keV count rates to 0.3--10 keV fluxes using the spectrum described in \ref{subsec:lc} and measured $R_{\rm Flux}$ by dividing the 2SXPS peak flux by these values. This threshold is shown in the black line in the comparison of the fluxes shown in Figure \ref{fig:xmmsl2comp}.

As XMMSL2's sensitivity limit is also relatively shallow, we used the same $R_{\rm Flux}\geq10$ as for 2RXS. This selected ten InactiveHost sources and one ActiveHost source as flaring.

\subsubsection{4XMM-DR9}

Finally, 4XMM-DR9 achieves significantly deeper depth than any of the other catalogues used here, including 2SXPS, with a sensitivity of $<10^{-14}$ erg cm$^{-2}$ s$^{-1}$ in the 0.2--12 keV. Thus, 4XMM-DR9 includes faint sources that have been missed by these other sources and includes $\sim550000$ sources detected during the period 3 February 2000 and 26 February 2019, again comparable to 2SXPS. However, its sky coverage is much smaller than the other catalogues, hence their inclusion in this comparison.

Unlike the other catalogues, 4XMM-DR9 is not matched to in the 2SXPS pipeline\footnote{4XMM-DR9 was released more recently than 2SXPS and 2SXPS is therefore matched to 3XMM-DR8 instead.}. We therefore performed our own cross-match, assuming a match if the separation was less than the combined 3-$\sigma$ position error. This produced 2908 (23.6\%) InactiveHost matches and 2719 (15.3\%) ActiveHost matches. For sources without either XMMSL2 or 4XMM-DR9 matches, we acquired upper limits from the the LEicester Database and Archive Service (LEDAS) using the \texttt{FLIX}\footnote{\url{https://www.ledas.ac.uk/flix/flix.html}} tool. \texttt{FLIX} is based on 3XMM-DR7 and therefore covers \textit{XMM-Newton} observations between 3 February 2000 and 15 December 2016, again roughly contemporary to the observations that make up 2SXPS.

The same power law spectrum was used to convert the 0.2--12 keV band count rates to 0.3--10 keV fluxes. However, a direct comparison to the 2SXPS peak fluxes was unsuitable due to the poorer sensitivity of 2SXPS, which led to significant Eddington bias which could obscure real flaring behaviour. Rather than a direct comparison of the fluxes and an $R_{\rm Flux}$ threshold, we fitted the full distribution of fluxes with a broken power law using the \texttt{emcee} Python module \citep{emcee}. Sources with 2SXPS peak fluxes greater than the fit by at least ten times their error were deemed to be flaring. This criterion was designed to minimise the risk of contamination due to Eddington bias and that only real outliers would be selected. This threshold indicated 224 InactiveHost and 309 ActiveHost sources were flaring.

\begin{figure}
	\includegraphics[width=\columnwidth]{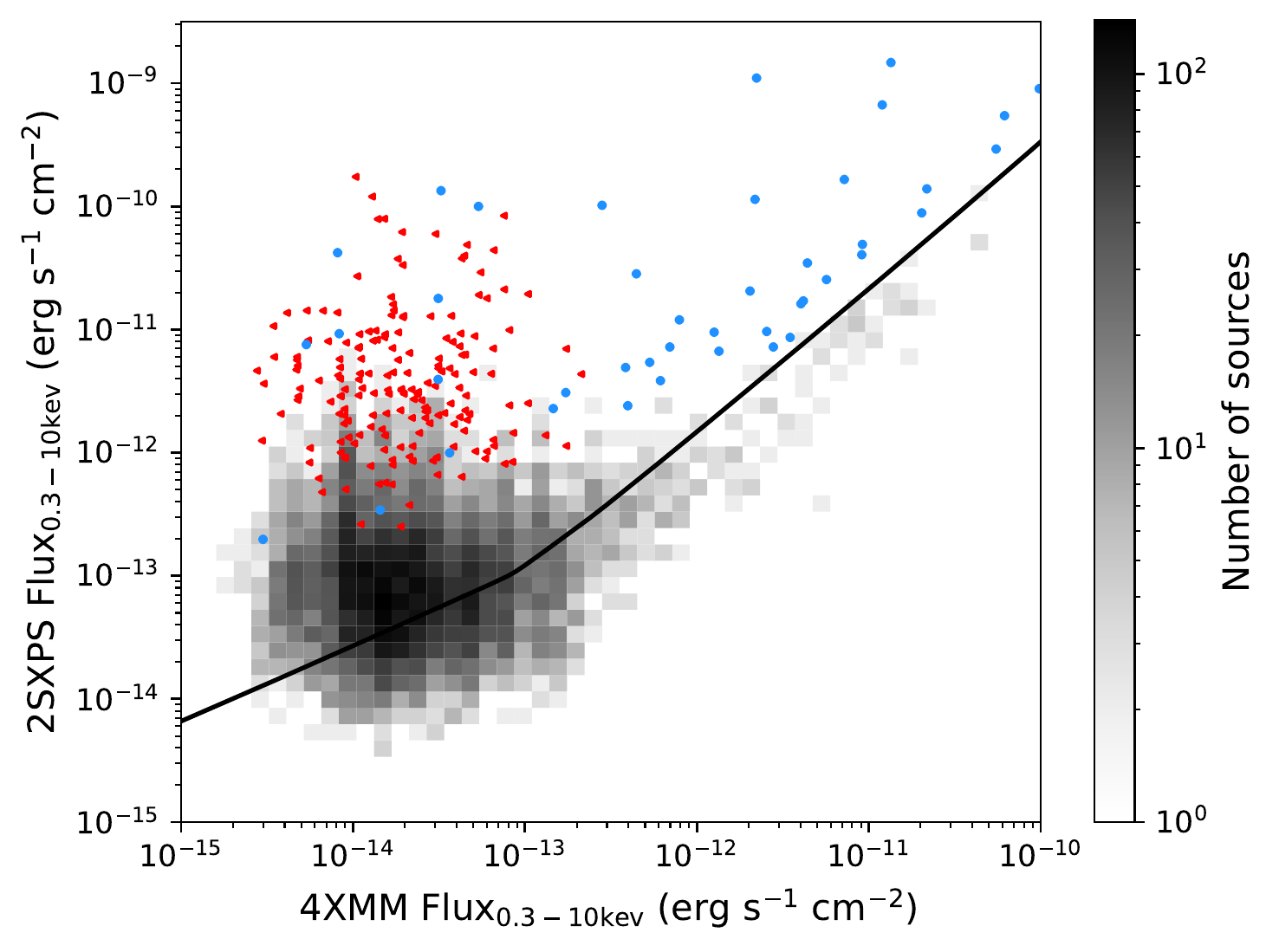}
	\includegraphics[width=\columnwidth]{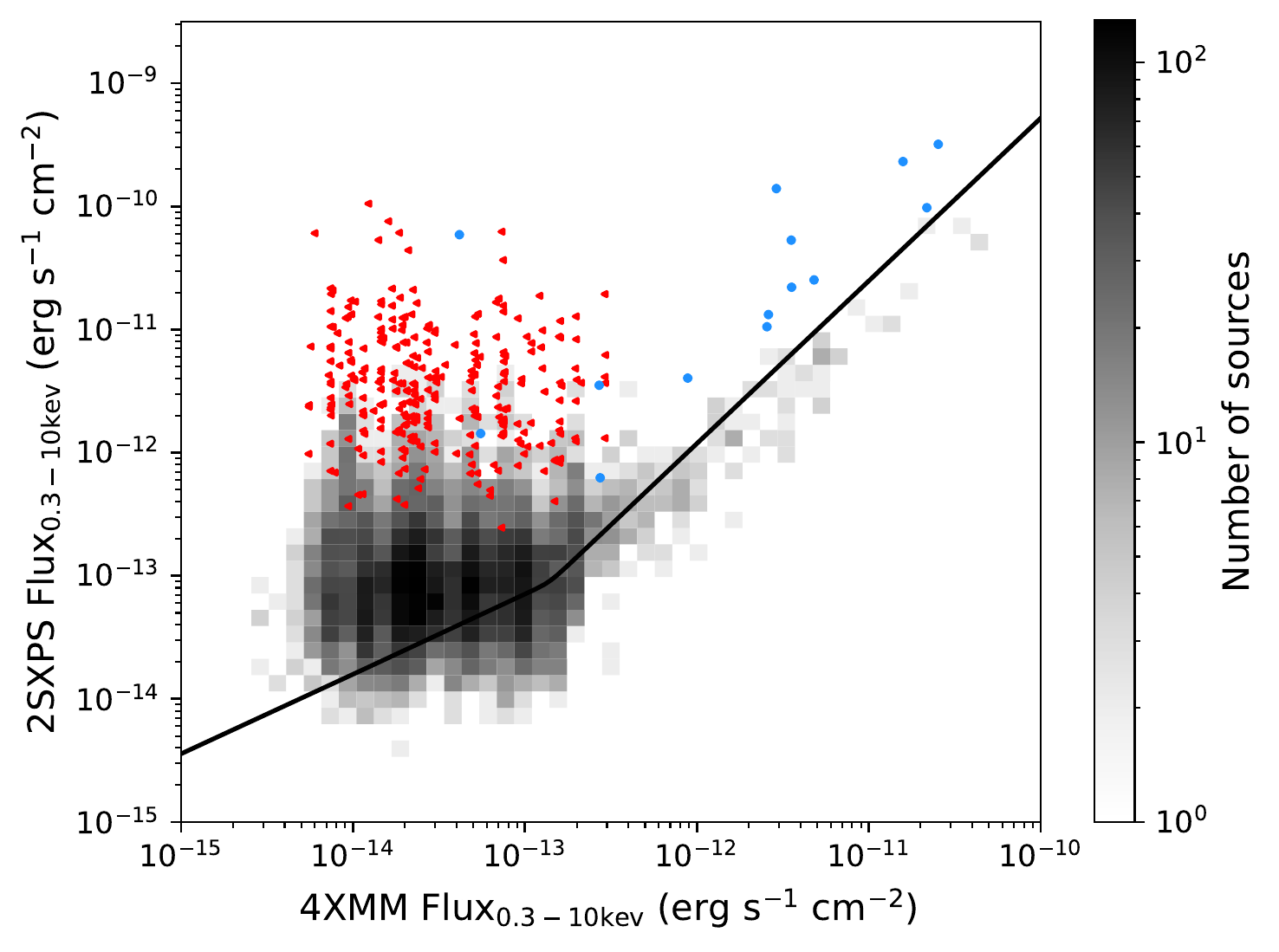}
    \caption{The flux of InactiveHost sources, top, and ActiveHost sources, bottom, in the 2SXPS and 4XMM-DR9 catalogues. The black line indicates the broken power law fit, the greyscale is the number density of the majority of the source population while sources matching the selection criterion are shown with the blue points indicating sources detected in both catalogues and the red triangles indicating sources with upper limits for derived using \texttt{FLIX}.}
    \label{fig:4xmmcomp}
\end{figure}

\begin{figure}
	\includegraphics[width=\columnwidth]{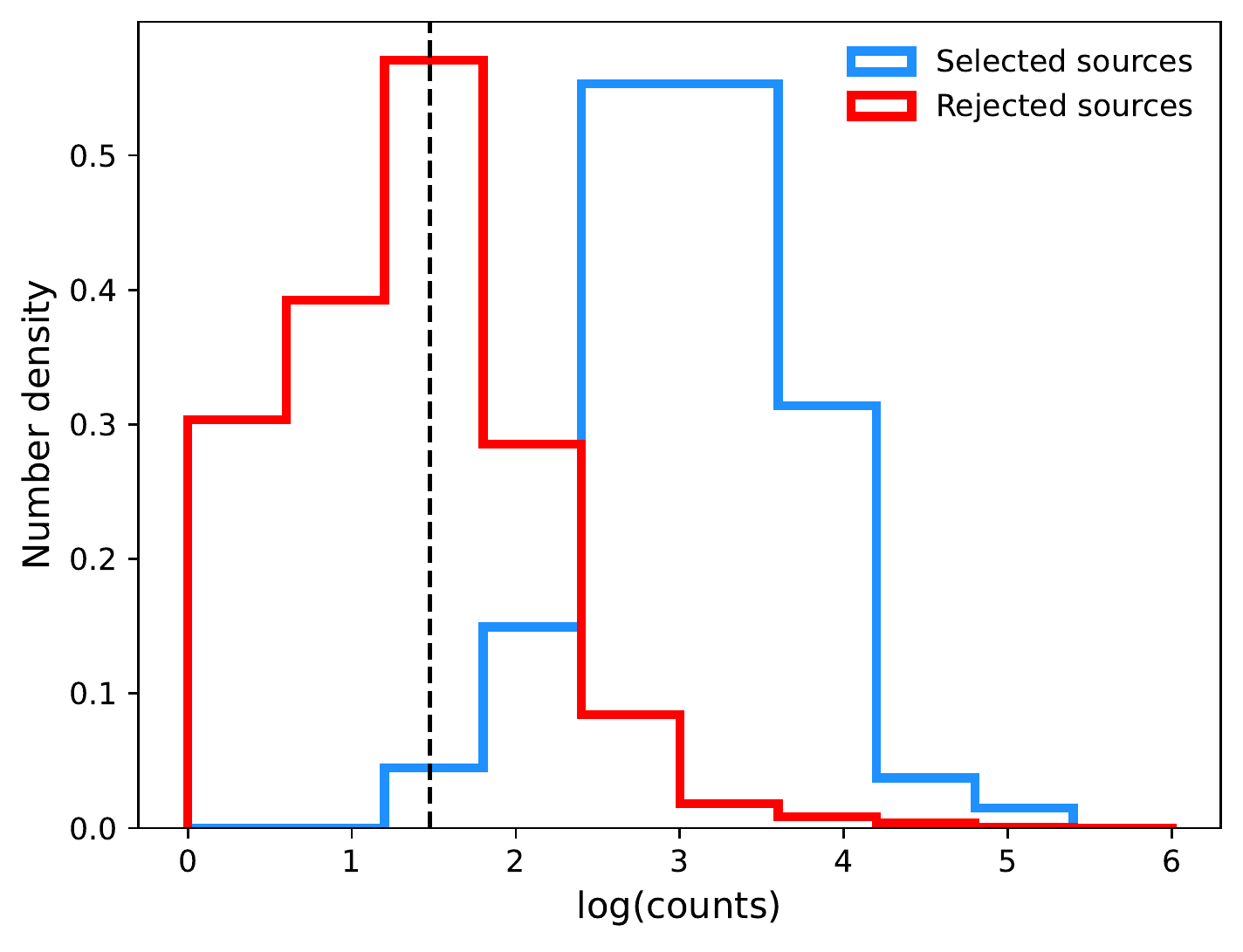}
	\includegraphics[width=\columnwidth]{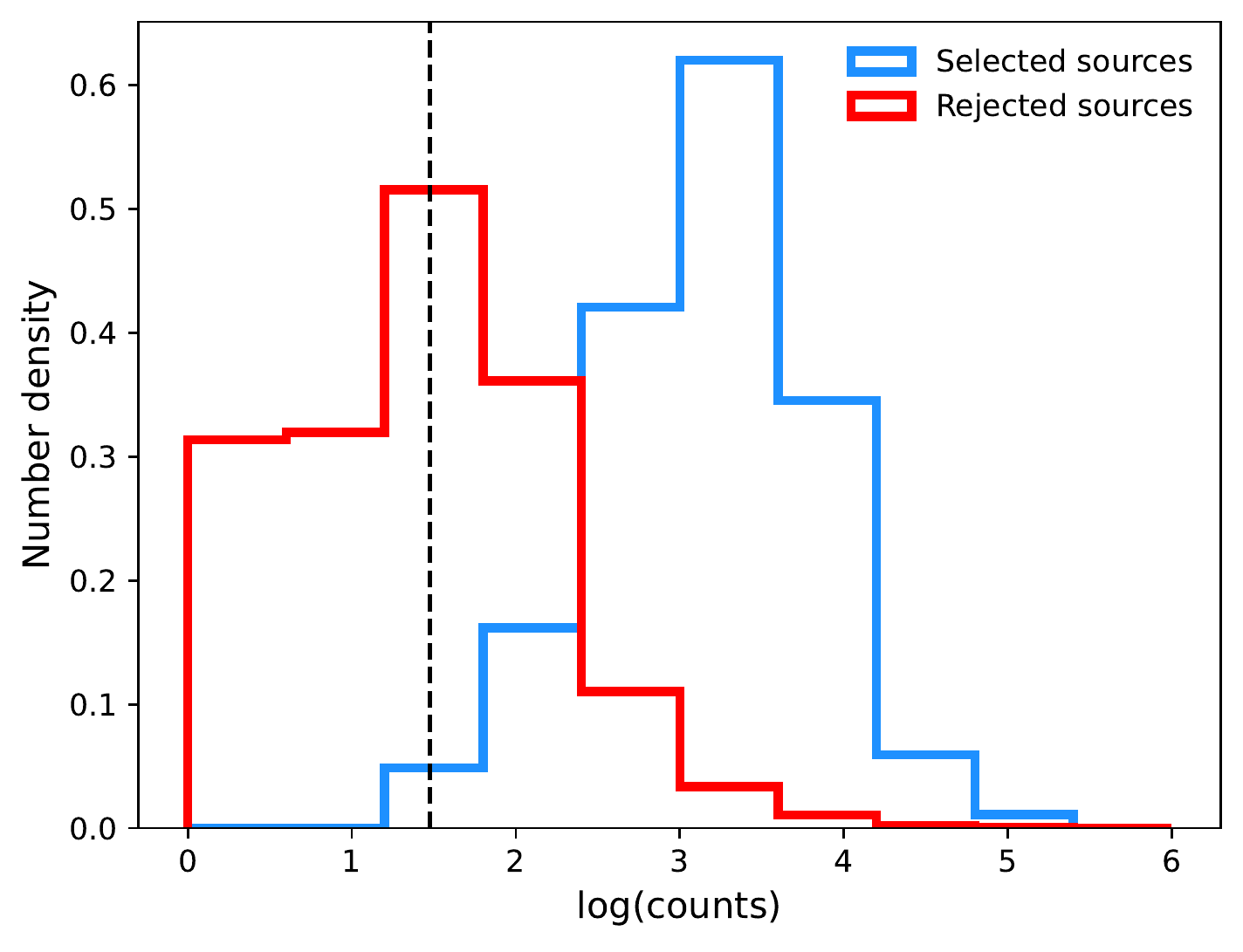}
    \caption{The distribution of counts in the peak 2SXPS epoch of InactiveHost sources, top, and ActiveHost sources, bottom. The blue histogram indicates sources selected by our criterion, the red histogram is those rejected and the black dashed line indicates the 30 counts, the value at which Eddington bias is likely to be negligible. Note that the y-axis indicates number density and the histogram heights should not be directly compared.}
    \label{fig:4xmmcounts}
\end{figure}

As an additional check, we also inspected the distributions of the number of counts detected for each source in their peak 2SXPS epoch. In the case of 1SXPS, Eddington bias was found to have less impact as the number of counts increased, with its significance greatly reduced at $>30$ counts \citep[see Figure 10 of][]{Evans14}. These distributions, shown in Figure \ref{fig:4xmmcounts}, indicate that our selection criterion does filter out the majority of the sources that are most likely to be affected by Eddington bias, although a fraction of them remain in the sample. We do not eliminate these sources at this stage, however, to prevent real transients being rejected as the Eddington bias was found to vary across the source population by \citet{Evans14}.

\subsection{Results}

As would be expected, there was some crossover in the sources selected using each criterion and a source was deemed to be flaring if it met at least one criterion. Overall, however, the selection criteria yielded 334 ($\sim2.6$\%) InactiveHost and 372 ($\sim2.1$\%) ActiveHost flaring sources. The majority of these sources were selected by the comparison to 4XMM-DR9/\texttt{FLIX}, and our analysis suggests that this could still be influenced by Eddington bias.

\section{Source classification}
\label{sec:src_class}

To perform an initial classification for each source, its 2SXPS light curve was manually examined in relation to the results of Section \ref{sec:flare_id}; to its position in the Set of Identifications, Measurements and Bibliography for Astronomical Data \citep[SIMBAD,][]{Wenger00} database\footnote{\url{https://simbad.u-strasbg.fr/simbad/}}; and to its webpage in the 2SXPS catalogue, which provides additional information such as pileup correction factors. SIMBAD had not been queried at an earlier stage due to the limit on the rate it could be queried to protect the servers. Each source was classified into one of several broad categories: spurious flares, Galactic sources, AGN activity, previously known transients and new transient candidates. To reduce AGN contamination, we made the assumption that the majority of X-ray activity in ActiveHost galaxies, or InactiveHost galaxies found to be active in SIMBAD, would be due to the AGN. Therefore, only extreme outliers in these hosts were selected as transient candidates. We were also stricter on flaring sources identified in the 4XMM-DR9/\texttt{FLIX} comparison to minimise the impact of Eddington bias. The results of this initial classification are given in Table \ref{tab:initialclass} and discussed further below.

\begin{table}
    \centering
        \caption{The classification of flaring sources in the InactiveHost and ActiveHost samples.}
    \begin{tabular}{ccc}
    \hline
       Category  & \multicolumn{2}{c}{\small Number (percentage of flaring sources)} \\
       & InactiveHost & ActiveHost \\
         \hline
       Spurious flares & 90 (26.9\%) & 28 (7.5\%)\\
       Galactic sources & 4 (1.2\%) & 0 (0.0\%) \\
       AGN & 93 (27.8\%) & 305 (82.0\%) \\
       Previously known transients & 138 (41.3\%) & 29 (7.8\%) \\
       New transient candidates & 9 (2.7\%) & 10 (2.7\%) \\
       \hline
    \end{tabular}
    \label{tab:initialclass}
\end{table}

\subsection{Spurious flares}

A significant proportion of flares in each sample were found to be spurious. Here, we examine some of the reasons why these sources were misidentified.

\subsubsection{Pileup}

Pileup can be a significant problem in some X-ray observations, particularly for bright sources. It arises when more than one photon arrives at the same CCD location within the readout time causing these photons to be registered as a single high energy photon or rejected as invalid detections \citep[e.g.][]{Ballet99,Davis01}.

For the 2SXPS catalogue, pileup is expected to become significant at count rates $\geq0.6$ counts s$^{-1}$, although the stochastic nature of photon arrivals means all sources are likely affected to some degree. The 2SXPS pipeline is designed to account for pileup via fitting the PSF of each source twice, once with a simple PSF model consisting of a Gaussian and a King component and again with an additional loss function component. Significant differences between these fits indicate pileup and the measured count rates adjusted to account for this \citep{Evans19}.

\begin{figure}
    \centering
    \includegraphics[width=\columnwidth]{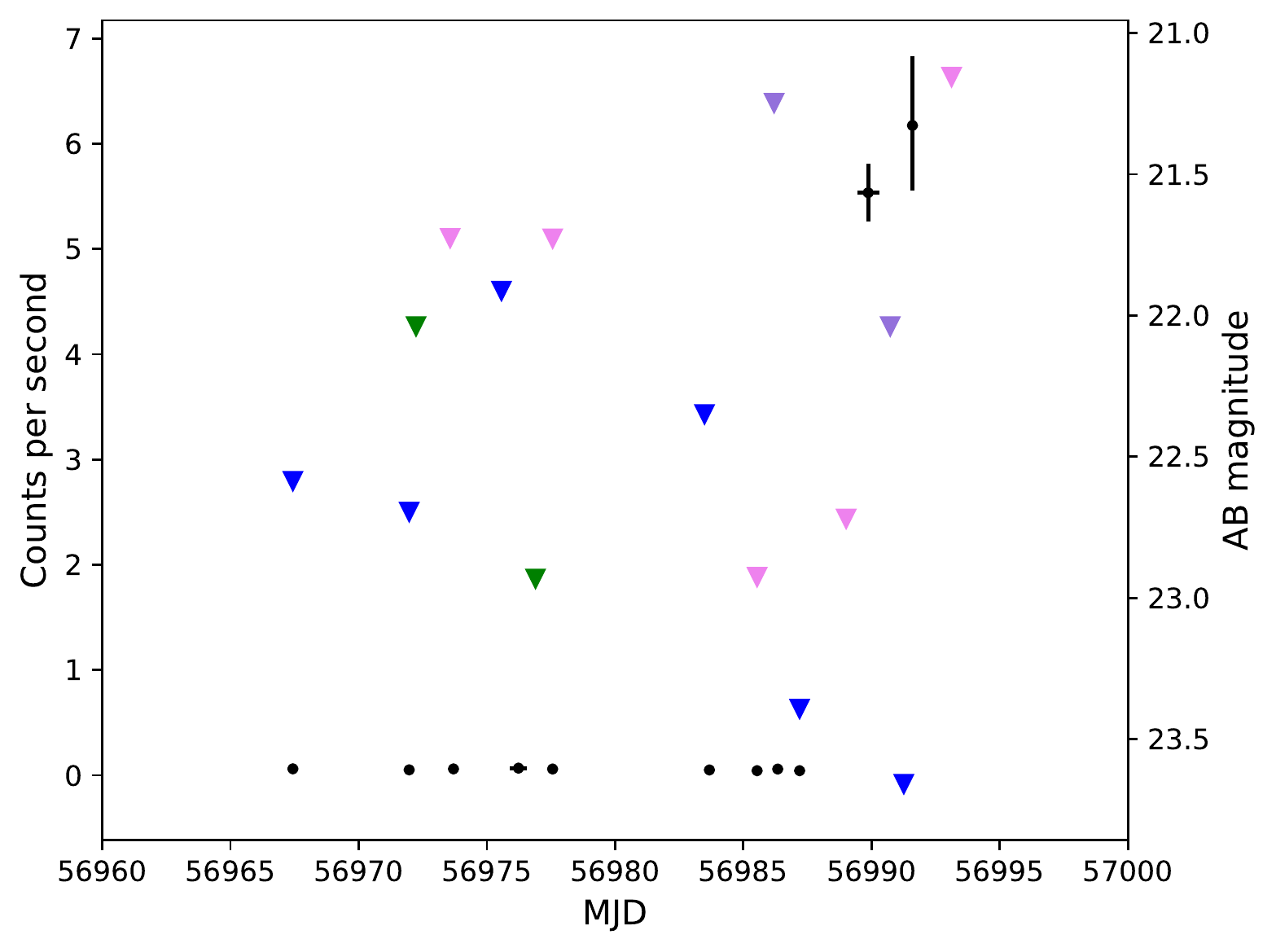}
    \caption[]{The portion of the XRT and UVOT light curve of 2SXPS J041128.2-634108 covering 30 October to 9 December 2014 showing the spurious flare caused by inaccurate pileup corrections.}
    \label{fig:J0411lc}
\end{figure}

However, if the pileup is inaccurately fitted, the applied correction factor can cause apparent flaring. For example, 2SXPS J041128.2-634108 was selected on the basis of two detections in late November 2014 that appeared much brighter than other detections of the source, as shown in Figure \ref{fig:J0411lc}. However, it was found that these detections had correction factors, dominated by inaccurately fitted pileup, two orders of magnitude larger than applied to other detections.

This behaviour is also displayed by a small number of other sources. It is not entirely clear what has led to this, but is plausibly due to models designed for point sources being applied to extended sources. To eliminate such sources, the correction factors were compared across the source's light curve and sources were rejected if the correction factor near the peak was significantly greater ($\sim$ one order of magnitude or more) but the uncorrected count rate was consistent with other detections. It should be noted that our methodology is designed to be sensitive to this kind of discrepancy and it is likely that inaccurate pileup is negligible for the vast majority of sources. It should also be noted that pileup has been further addressed for LSXPS and the inaccurate corrections have been eliminated for 2SXPS J041128.2-634108, for instance.

\subsubsection{Eddington bias}

A number of sources, particularly in the ActiveHost sample, were rejected due to the likelihood that the apparent flaring behaviour was actually due to Eddington bias in the 2SXPS detections. These sources were selected based on the fact that they were only deemed to be flaring by the 4XMM-DR9/\texttt{FLIX} criterion and all other criteria were consistent with the source being in a steady state.

\subsubsection{Minor factors}

A small number of sources were rejected as they appeared to be spurious detections, possibly due to high separation between the pointing centre and the `source'.
For example, both 2SXPS J111936.2+134044 and 2SXPS J111932.2+133853 were only detected in one observation in a band of similar features that were deemed noise by the 2SXPS pipeline.

Finally, there is a possibility that some sources were incorrectly matched to the external catalogues and their properties were therefore incorrectly assumed. In particular, in XMMSL2, there is a long tail in the positional error which could affect a significant number of sources. However, Figures \ref{fig:2rxscomp}, \ref{fig:xmmsl2comp} and \ref{fig:4xmmcomp} are broadly consistent with a slope of unity, indicating that any incorrect cross matching of this kind is limited to a very small sample of sources.

\subsection{Galactic sources}

Galactic sources were identified using the SIMBAD query. Our method proved successful at minimising contamination from such sources with only four across both samples appearing to be Galactic.

\subsection{AGN}

As would be expected, the largest contribution to flaring sources in the ActiveHost sample was AGN activity. However, we also found AGN to make a significant contribution to the InactiveHost sample. This is likely due to the incompleteness of the AGN catalogues we used to check for activity. However, SIMBAD draws on many other data sources on top of these catalogues and is therefore significantly more complete. Ideally, the activity check would therefore be performed by querying SIMBAD but this was not possible due to the querying limit mentioned above.

\subsection{Previously known transients}

2SXPS also includes a significant number of previously known transients. In situations where it was unclear whether a flare was a new or an historical transient, the swiftmastr catalogue\footnote{\url{https://www.swift.ac.uk/swift_live/index.php}} was queried to determine the target of the observation. If the 2SXPS source was the main target, it was assumed to be a previously known transient.

\section{Candidate characterisation}
\label{sec:transient_character}

From our analysis, 19 possible new transient candidates were identified, as summarised in Table \ref{tab:candidates}. These sources were defined as faint or bright depending on whether enough photons ($\sim50$ background subtracted counts) had been detected to accurately fit a spectrum. In this Section, we discuss the properties of this sample of new transient candidates.

\subsection{UVOT data}

To more fully investigate the transients, we retrieved the associated \textit{Swift}-UVOT data from the UKSSDC, selecting all observations within 12 arcminutes of the 2SXPS sources' positions. These data had been fully processed by the automatic pipeline and sources were extracted from the data using the \texttt{uvotdetect} and \texttt{uvotsource} utilities \citep{Poole08,Immler08,Breeveld10,heasoft} with a 5-$\sigma$ threshold. If no source was detected consistent with the 2SXPS position, \texttt{uvotsource} was rerun with a 3-$\sigma$ threshold. If a source was detected with this threshold, it was used to define an upper limit, otherwise the upper limit was taken to be the limiting magnitude of the exposure. We found that five sources had UVOT counterparts and identified upper limits for all but two of the remaining sources (2SXPS J215459.6-154104 and 2SXPS J001319.7-600638). It should be noted that the smaller UVOT field of view compared to the XRT ($17\times17$ arcmin versus $23.6\times23.6$ arcmin) also contributed to the lack of UVOT counterparts in some cases.

\begin{table*}
    \centering
        \caption{The final sample of new transient candidates identified in 2SXPS. Starred redshifts, and therefore those sources' luminosities, are assumed based on the InactiveHost or ActiveHost sample's mean. Bright/Faint sources are defined as in the text. Bright sources' luminosities are calculated from the best fitting spectrum (see Section \ref{sec:bright}).}
    \renewcommand{\arraystretch}{1.5}
    \begin{tabular}{cccccccc}
    \hline
        2SXPS ID &  2SXPS name &  Host name &  Active &  $z$ &  $L_{\rm pk}$ &  Bright/Faint  &  UVOT \\
      & & & host? & & (erg s$^{-1}$) &  &  detected? \\
     \hline    
 5396 & 2SXPS J123320.7+210524 & SDSS J123320.67+210529.3 & Y & 1.082* & $4.07^{+2.30}_{-1.75}\times 10^{45}$ & F & N \\
27717 & 2SXPS J074553.0+254444 & WISEA J074553.14+254444.0 & Y & 1.082* & $7.30 {\footnotesize \pm} 0.53 \times 10^{45}$ & B & Y \\
56527 & 2SXPS J102526.2+170847 & WISEA J102525.83+170842.4 & Y & 1.082* & $1.48^{+0.54}_{-0.44}\times 10^{45}$ & F & N \\
92008 & 2SXPS J052627.4-211713 & ESO 553- G 043 & Y & 0.028 & $4.27 {\footnotesize \pm} 0.30 \times 10^{42}$ & B & Y \\
94518 & 2SXPS J123340.7+205747 & SDSS J123340.92+205735.8 & N & 0.326* & $1.68^{+1.35}_{-1.03}\times 10^{44}$ & F & N \\
100147 & 2SXPS J140456.5+542352 & SDSS J140456.38+542353.9 & N & 0.326* & $1.39^{+2.18}_{-1.06}\times 10^{45}$ & F & N \\
106084 & 2SXPS J020751.5+445038 & UGC 01609 & N & 0.022 &  $8.17  {\footnotesize \pm}  0.49 \times 10^{42}$ & B & Y \\
109927 & 2SXPS J215459.6-154104 & WISEA J215459.79-154104.5 & Y & 1.082* &  $9.87  {\footnotesize \pm}  0.91 \times 10^{45}$ & B & N \\
118094 & 2SXPS J114928.2+082804 & SDSS J114928.19+082759.3 & N & 0.326* & $5.80^{+0.95}_{-0.85}\times 10^{44}$ & F & Y \\
127404 & 2SXPS J001319.7-600638 & WISEA J001319.35-600642.5 & Y & 1.082* &  $2.83  {\footnotesize \pm}  0.25 \times 10^{45}$ & B & N \\
127981 & 2SXPS J160958.1+181733 & WISEA J160957.79+181735.1 & N & 0.326* & $7.39^{+7.26}_{-4.45}\times 10^{43}$ & F & N \\
139825 & 2SXPS J020444.6+004814 & WISEA J020444.50+004812.2 & N & 0.326* & $1.25^{+1.18}_{-0.72}\times 10^{44}$ & F & N \\
146218 & 2SXPS J151944.7+234807 & SDSS J151943.90+234802.1 & Y & 1.082* & $2.20^{+1.67}_{-1.12}\times 10^{45}$ & F & N \\
153522 & 2SXPS J014831.1+273213 & WISEA J014831.20+273211.5 & N & 0.477 & $3.46^{+1.88}_{-1.39}\times 10^{44}$ & F & N \\
162692 & 2SXPS J062148.6+744200 & WISEA J062148.90+744204.5 & Y & 1.082* &  $2.18  {\footnotesize \pm}  0.11 \times 10^{47}$ & B & N \\
179179 & 2SXPS J191909.8+440937 & CGCG 229-037 & N & 0.040 &  $4.85  {\footnotesize \pm}  0.24 \times 10^{43}$ & B & Y \\
192646 & 2SXPS J054919.2-620513 & [BRS2016] J054919.69-620507.7 & N & 0.377 &  $2.40  {\footnotesize \pm}  0.21 \times 10^{47}$ & B & N \\
196450 & 2SXPS J163703.0-495137 & XTE J1637-498 & Y & 1.082* &  $2.32  {\footnotesize \pm}  0.14 \times 10^{47}$ & B & N \\
206310 & 2SXPS J132507.2-323813 & WISEA J132507.02-323812.4 & Y & 1.082* & $2.12^{+0.67}_{-0.55}\times 10^{46}$ & F & N \\
    \hline
    \end{tabular}
    \label{tab:candidates}
\end{table*}

\subsection{Faint sources}

The majority of the transient candidate sample, 10 out of the 19 sources, had not accumulated sufficient counts to have accurate spectra fitted. Their low count rates indicate that this is due to them generally being inherently faint sources rather than a product of shorter exposure times and are labelled as `Faint' in Table \ref{tab:candidates}. This also contributed to a lack of UVOT candidates across this subsample. Unfortunately, due to the low number of accumulated counts, these sources could not be characterised in much depth. However, their 2SXPS light curves are presented in Figure \ref{fig:faintlc}.

\subsection{Bright sources}
\label{sec:bright}

The nine remaining sources had accumulated enough counts for spectra to be accurately fitted and are designated as `Bright' sources in Table \ref{tab:candidates}. Their 2SXPS light curves are shown in Figure \ref{fig:brightlc}.

\subsubsection{Integrated spectra}

As part of the 2SXPS pipeline, such sources are automatically fitted with both power law and Astrophysical Plasma Emission Code (APEC) \citep{Smith01} models. Power laws are representative of AGN or several classes of transient such as gamma-ray bursts (GRBs), while APECs represent shock heated plasmas such as those in dark matter haloes or the inner regions of black hole accretion disks. However, the pipeline assumes a redshift of $z=0$ and we therefore repeated these fits using \texttt{XSpec 12.10.0} \citep{Arnaud96} and our redshifts given in Table \ref{tab:candidates}. In addition to the power law and APEC models, we also fitted the transient candidates with a blackbody model, typical for other transients such as TDEs. All our spectral fits included both host and Milky Way extinction using \texttt{tbabs} and \texttt{ztbabs} components \citep{XSPECtbabs}. These spectral fits are summarised in Tables \ref{tab:plaw}, \ref{tab:apec} and \ref{tab:bb}.

\begin{table*}
    \centering
        \caption{The spectral fits of the bright transient candidates using a power law model. $N_{H, \text{Gal}}$ is taken from \citet{Willingale13} and all $N_H$ values are given in units of 10$^{21}$ cm$^{-2}$. Starred values of $N_{H, \text{host}}$ indicate sources where the fit was insensitive to this parameter.}
    \begin{tabular}{ccccccc}
    \hline
       2SXPS ID  & 2SXPS name & $N_{H, \text{Gal}}$ & $\Gamma$ & $N_{H, \text{host}}$ & $L_{\rm pk, 0.3-10 keV}$ (erg s$^{-1}$) & Red. $\chi^2$ \\
    \hline
27717 & 2SXPS J074553.0+254444 & 0.54 & 1.83$^{+0.15}_{-0.09}$ & $<1.13$ & $7.30 \pm 0.53 \times 10^{45}$ & 1.06  \\
92008 & 2SXPS J052627.4-211713 & 0.33 & 1.19$^{+0.69}_{-0.60}$ & 150.11$^{+55.51}_{-42.87}$ & $4.27 \pm 0.30 \times 10^{42}$ & 0.93  \\
106084 & 2SXPS J020751.5+445038 & 1.21 & 1.69$^{+0.17}_{-0.16}$ & 4.87$^{+1.27}_{-1.12}$ & $8.17 \pm 0.49 \times 10^{42}$ & 1.19  \\
109927 & 2SXPS J215459.6-154104 & 0.43 & 1.66$^{+0.26}_{-0.26}$ & $<0.85$ & $1.33 \pm 0.12 \times 10^{46}$ & 0.84  \\
127404 & 2SXPS J001319.7-600638 & 0.14 & 1.97$^{+0.39}_{-0.26}$ & 0.47$^{+2.43}_{-0.47}$ & $6.73 \pm 0.60 \times 10^{45}$ & 0.72  \\
162692 & 2SXPS J062148.6+744200 & 1.19 & 1.72$^{+0.19}_{-0.16}$ & 0.77$^{+2.22}_{-0.77}$ & $2.18 \pm 0.11 \times 10^{47}$ & 1.01  \\
179179 & 2SXPS J191909.8+440937 & 1.09 & 1.73$^{+0.13}_{-0.12}$ & $<0.17$ & $4.85 \pm 0.24 \times 10^{43}$ & 0.94  \\
192646 & 2SXPS J054919.2-620513 & 0.53 & 1.54$^{+0.31}_{-0.28}$ & 0.62$^{+1.38}_{-0.62}$ & $2.40 \pm 0.21 \times 10^{47}$ & 0.84  \\
196450 & 2SXPS J163703.0-495137 & 12.10 & 1.10$^{+0.42}_{-0.33}$ & 15.57$^{+42.06}_{-15.57}$ & $2.31 \pm 0.14 \times 10^{47}$ & 1.09  \\
         \hline
    \end{tabular}
    \label{tab:plaw}
\end{table*}

\begin{table*}
    \centering
        \caption{The spectral fits of the bright transient candidates using an APEC model. $N_{H, \text{Gal}}$ is taken from \citet{Willingale13} and all $N_H$ values are given in units of 10$^{21}$ cm$^{-2}$. Starred values of $N_{H, \text{host}}$ indicate sources where the fit was insensitive to this parameter and starred values of  k$T$ indicate sources where this parameter was poorly constrained.}
    \begin{tabular}{ccccccc}
    \hline
       2SXPS ID  & 2SXPS name & $N_{H, \text{Gal}}$ & k$T$ (keV) & $N_{H, \text{host}}$ & $L_{\rm pk, 0.3-10 keV}$ (erg s$^{-1}$) & Red. $\chi^2$ \\
    \hline
27717 & 2SXPS J074553.0+254444 & 0.54 & 8.56$^{+2.93}_{-1.62}$ & $<0.20$ & $4.46 \pm 0.32 \times 10^{45}$ & 1.06  \\
92008 & 2SXPS J052627.4-211713 & 0.33 & <64.00* & 157.23$^{+24.39}_{-21.81}$ & $2.17 \pm 0.15 \times 10^{43}$ & 2.57  \\
106084 & 2SXPS J020751.5+445038 & 1.21 & 16.02$^{+10.31}_{-5.91}$ & 3.19$^{+0.73}_{-0.65}$ & $8.33 \pm 0.50 \times 10^{42}$ & 1.32  \\
109927 & 2SXPS J215459.6-154104 & 0.43 & 23.60$^{+38.37}_{-13.97}$ & $<0.59$ & $9.87 \pm 0.91 \times 10^{45}$ & 1.11  \\
127404 & 2SXPS J001319.7-600638 & 0.14 & 5.20$^{+3.27}_{-1.48}$ & $<1.06$ & $3.48 \pm 0.31 \times 10^{45}$ & 0.71  \\
162692 & 2SXPS J062148.6+744200 & 1.19 & 18.47$^{+10.17}_{-6.13}$ & $<0.63$ & $1.60 \pm 8.20 \times 10^{47}$ & 1.10  \\
179179 & 2SXPS J191909.8+440937 & 1.09 & 7.52$^{+4.61}_{-1.98}$ & $<0.07$ & $3.54 \pm 0.17 \times 10^{43}$ & 1.14  \\
192646 & 2SXPS J054919.2-620513 & 0.53 & 8.09$^{+9.23}_{-2.93}$ & 0.44$^{+1.04}_{-0.44}$ & $1.64 \pm 0.14 \times 10^{47}$ & 0.83  \\
196450 & 2SXPS J163703.0-495137 & 12.10 & <63.91* & 34.35$^{+28.89}_{-19.38}$ & $2.32 \pm 0.14 \times 10^{47}$ & 1.06  \\
         \hline
    \end{tabular}
    \label{tab:apec}
\end{table*}

\begin{table*}
    \centering
        \caption{The spectral fits of the bright transient candidates using a blackbody model. $N_{H, \text{Gal}}$ is taken from \citet{Willingale13} and all $N_H$ values are given in units of 10$^{21}$ cm$^{-2}$. Starred values of $N_{H, \text{host}}$ indicate sources where the fit was insensitive to this parameter.}
    \begin{tabular}{ccccccc}
    \hline
    2SXPS ID  & 2SXPS name & $N_{H, \text{Gal}}$ & k$T$ (keV) & $N_{H, \text{host}}$ & $L_{\rm pk, 0.3-10 keV}$ (erg s$^{-1}$) & Red. $\chi^2$ \\
    \hline
27717 & 2SXPS J074553.0+254444 & 0.54 & 1.08$^{+0.07}_{-0.07}$ & $<0.11$ & $3.40 \pm 0.25 \times 10^{45}$ & 1.59  \\
92008 & 2SXPS J052627.4-211713 & 0.33 & 2.13$^{+0.59}_{-0.39}$ & 107.75$^{+36.61}_{-29.30}$ & $1.38 \pm 9.84 \times 10^{43}$ & 0.92  \\
106084 & 2SXPS J020751.5+445038 & 1.21 & 0.97$^{+0.04}_{-0.04}$ & $<0.22$ & $5.78 \pm 0.35 \times 10^{42}$ & 1.20  \\
109927 & 2SXPS J215459.6-154104 & 0.43 & 1.54$^{+0.28}_{-0.23}$ & $0.00$* & $7.94 \pm 0.73 \times 10^{45}$ & 1.68  \\
127404 & 2SXPS J001319.7-600638 & 0.14 & 0.86$^{+0.11}_{-0.10}$ & $<0.39$ & $2.83 \pm 0.25 \times 10^{45}$ & 0.76  \\
162692 & 2SXPS J062148.6+744200 & 1.19 & 1.39$^{+0.10}_{-0.09}$ & $<0.29$ & $1.17 \pm 6.02 \times 10^{47}$ & 1.27  \\
179179 & 2SXPS J191909.8+440937 & 1.09 & 0.68$^{+0.05}_{-0.04}$ & $0.00$* & $2.57 \pm 0.12 \times 10^{43}$ & 1.71  \\
192646 & 2SXPS J054919.2-620513 & 0.53 & 0.94$^{+0.09}_{-0.09}$ & $<0.23$ & $1.33 \pm 0.12 \times 10^{47}$ & 0.84  \\
196450 & 2SXPS J163703.0-495137 & 12.10 & 2.69$^{+0.31}_{-0.28}$ & $<10.81$ & $1.80 \pm 0.11 \times 10^{47}$ & 0.93  \\
         \hline
    \end{tabular}
    \label{tab:bb}
\end{table*}

From our reduced $\chi^2$ values, most candidates are reasonably well fit with all three models. However, there were some notable exceptions. In some cases, the plasma temperatures inferred from the APEC model fits were poorly constrained. For example, both 2SXPS 2SXPS J052627.4-211713 and 2SXPS J163703.0-495137 have fitted temperatures of $\sim$64 keV, the maximum allowed in \texttt{XSpec}'s APEC model, likely indicating the true temperature would be greater than this. This means the APEC model is most likely unsuitable for such sources, an expected result as the APEC model is designed for a relatively small subset of X-ray sources.

The fits for a significant proportion of the sources were also found to be insensitive to the host absorption. This, again, is not necessarily unexpected and likely indicates the host absorption is significantly smaller than that of the Milky Way.

Overall, of the nine sources, four are fit best with power laws, two with APECs and one with a blackbody. The exceptions are 2SXPS J074553.0+254444, which is equally well fitted by both the power law and APEC models, and 2SXPS J054919.2-620513, which is equally well fitted with power law and blackbody models. It should be noted that these fits are sensitive to redshift, however.

For the population as a whole, the distribution of photon index $\Gamma$ is consistent with several classes of X-ray source including AGN \citep[e.g.][]{Corral11}, GRBs \citep[e.g.][]{DAvanzo14,Fong15} and high mass X-ray binaries \citep[HMXB,][]{Sazonov17}. However, the APEC fits imply higher temperatures than those typically found in AGN \citep[e.g.][]{David11,Noda13} while the blackbody temperatures are also significantly higher than many transient classes such as TDEs \citep[e.g.][]{Dai15,Saxton20,Saxton21}. The majority of candidates in Table \ref{tab:candidates} are therefore unlikely to be thermally dominated.

\subsubsection{Spectral variance with time}

In addition to their full integrated spectra, four sources had accumulated enough counts in two or more individual observations for spectra to be fitted. Variations between epochs can also help to characterise such sources. We therefore repeated our procedure for fitting the spectra for each epoch with the power law, APEC and blackbody models. These fits are summarised in Tables \ref{tab:epochpl}, \ref{tab:epochapec} and \ref{tab:epochbb}.

\begin{table*}
    \centering
        \caption{The spectral fits of the individual epochs of the bright transient candidates using a power law model. Starred values of $N_{H, \text{host}}$ indicate sources where the fit was insensitive to this parameter.
        }
    \begin{tabular}{cccccc}
    \hline
        Epoch  &  MJD & $\Gamma$ &  $N_{H, \text{host}}$ (10$^{21}$ cm$^{-2}$) & $L_{\rm 0.3-10 keV}$ (erg s$^{-1}$) &  Red. $\chi^2$ \\
    \hline
\multicolumn{6}{c}{2SXPS J074553.0+254444} \\
1 & 53678.03 & 1.95$^{+0.29}_{-0.21}$ & 0.05$^{+0.22}_{-0.05}$ & $4.69 \pm 0.34 \times 10^{45}$ & 0.76 \\
2 & 53679.04 & 1.85$^{+0.16}_{-0.16}$ & $<0.75$ & $4.76 \pm 0.35 \times 10^{45}$ & 1.07 \\
3 & 53737.09 & 1.67$^{+0.40}_{-0.26}$ & $<3.23$ & $2.71 \pm 0.31 \times 10^{45}$ & 1.05 \\
\hline
\multicolumn{6}{c}{2SXPS J052627.4-211713} \\
1 & 55332.02 & 1.72$^{+1.12}_{-0.98}$ & 18.69$^{+8.80}_{-6.90}$ & $2.25 \pm 0.16 \times 10^{43}$ & 1.30 \\
2 & 57616.52 & 0.52$^{+0.97}_{-0.79}$ & 11.97$^{+8.47}_{-5.38}$ & $1.73 \pm 0.12 \times 10^{43}$ & 1.25 \\
\hline
\multicolumn{6}{c}{2SXPS J020751.5+445038} \\
1 & 55489.38 & 1.25$^{+0.74}_{-0.59}$ & 0.21$^{+0.55}_{-0.21}$ & $5.64 \pm 1.37 \times 10^{42}$ & 1.12 \\
2 & 55631.12 & 1.59$^{+0.49}_{-0.43}$ & 0.46$^{+0.41}_{-0.28}$ & $8.94 \pm 0.79 \times 10^{42}$ & 0.82 \\
3 & 55647.52 & 1.99$^{+0.41}_{-0.37}$ & 0.83$^{+0.38}_{-0.30}$ & $1.30 \pm 0.07 \times 10^{43}$ & 0.83 \\
4 & 57333.82 & 1.66$^{+0.28}_{-0.25}$ & 0.43$^{+0.20}_{-0.16}$ & $8.91 \pm 0.44 \times 10^{42}$ & 1.26 \\
5 & 57334.02 & 1.51$^{+0.56}_{-0.49}$ & 0.19$^{+0.35}_{-0.19}$ & $4.83 \pm 0.55 \times 10^{42}$ & 0.94 \\
\hline
\multicolumn{6}{c}{2SXPS J215459.6-154104} \\
1 & 54314.06 & 1.71$^{+0.30}_{-0.29}$ & $<0.99$ & $9.12 \pm 0.84 \times 10^{45}$ & 0.85 \\
2 & 54315.0 & 2.49$^{+1.49}_{-0.93}$ & $<6.96$ & $5.74 \pm 0.46 \times 10^{45}$ & 0.61 \\
\hline
    \end{tabular}
    \label{tab:epochpl}
\end{table*}

\begin{table*}
    \centering
        \caption{The spectral fits of the individual epochs of the bright transient candidates using an APEC model. Starred values of $N_{H, \text{host}}$ indicate sources where the fit was insensitive to this parameter and starred values of  k$T$ indicate sources where this parameter was poorly constrained.}
    \begin{tabular}{cccccc}
    \hline
        Epoch  &  MJD & k$T$ (keV) &  $N_{H, \text{host}}$ (10$^{21}$ cm$^{-2}$) & $L_{\rm 0.3-10 keV}$ (erg s$^{-1}$) &  Red. $\chi^2$ \\
    \hline
\multicolumn{6}{c}{2SXPS J074553.0+254444} \\
1 & 53678.03 & 8.23$^{+6.82}_{-3.02}$ & $<0.61$ & $4.46 \pm 0.32 \times 10^{45}$ & 0.86 \\
2 & 53679.04 & 8.05$^{+4.06}_{-2.45}$ & $<0.32$ & $4.34 \pm 0.31 \times 10^{45}$ & 1.11 \\
3 & 53737.09 & 15.80$^{+35.06}_{-9.26}$ & $<1.45$ & $2.63 \pm 0.30 \times 10^{45}$ & 1.05 \\
\hline
\multicolumn{6}{c}{2SXPS J052627.4-211713} \\
1 & 55332.02 & 9.75$^{+-29.04}_{-19.30}$ & 15.67$^{+4.22}_{-2.70}$ & $1.70 \pm 0.12 \times 10^{43}$ & 1.32 \\
2 & 57616.52 & 17.54$^{+2.59}_{-3.54}$ & 17.54$^{+2.59}_{-3.54}$ & $2.38 \pm 0.17 \times 10^{43}$ & 1.38 \\
\hline
\multicolumn{6}{c}{2SXPS J020751.5+445038} \\
1 & 55489.38 & $<28.33$* & 0.25$^{+0.37}_{-0.23}$ & $5.44 \pm 1.32 \times 10^{42}$ & 1.14 \\
2 & 55631.12 & $<25.27$* & 0.32$^{+0.22}_{-0.18}$ & $8.24 \pm 0.73 \times 10^{42}$ & 0.84 \\
3 & 55647.52 & 10.59$^{+17.37}_{-6.05}$ & 0.49$^{+0.31}_{-0.16}$ & $9.81 \pm 0.59 \times 10^{42}$ & 0.92 \\
4 & 57333.82 & 11.82$^{+43.29}_{-5.52}$ & 0.32$^{+0.13}_{-0.13}$ & $8.04 \pm 0.40 \times 10^{42}$ & 1.33 \\
5 & 57334.02 & $<18.81$* & 0.12$^{+0.27}_{-0.12}$ & $4.55 \pm 0.52 \times 10^{42}$ & 0.95 \\
\hline
\multicolumn{6}{c}{2SXPS J215459.6-154104} \\
1 & 54314.06 & $<17.32$* & $<0.71$ & $9.31 \pm 0.86 \times 10^{45}$ & 1.08 \\
2 & 54315.0 & 2.78$^{+19.50}_{-1.35}$ & $<5.35$ & $5.06 \pm 0.40 \times 10^{45}$ & 0.62 \\
\hline
    \end{tabular}
    \label{tab:epochapec}
\end{table*}

\begin{table*}
    \centering
        \caption{The spectral fits of the individual epochs of the bright transient candidates using a blackbody model. Starred values of $N_{H, \text{host}}$ indicate sources where the fit was insensitive to this parameter.}
    \begin{tabular}{cccccc}
    \hline
        Epoch  &  MJD & k$T$ (keV) &  $N_{H, \text{host}}$ (10$^{21}$ cm$^{-2}$) & $L_{\rm 0.3-10 keV}$ (erg s$^{-1}$) &  Red. $\chi^2$ \\
    \hline
\multicolumn{6}{c}{2SXPS J074553.0+254444} \\
1 & 53678.03 & 0.96$^{+0.13}_{-0.13}$ & $0.00$* & $3.20 \pm 0.23 \times 10^{45}$ & 1.22 \\
2 & 53679.04 & 1.12$^{+0.12}_{-0.11}$ & $0.00$* & $3.47 \pm 0.25 \times 10^{45}$ & 2.12 \\
3 & 53737.09 & 1.10$^{+0.22}_{-0.18}$ & $0.00$* & $1.80 \pm 0.20 \times 10^{45}$ & 1.20 \\
\hline
\multicolumn{6}{c}{2SXPS J052627.4-211713} \\
1 & 55332.02 & 1.81$^{+0.75}_{-0.43}$ & 12.99$^{+6.08}_{-4.69}$ & $1.04 \pm 0.07 \times 10^{43}$ & 1.31 \\
2 & 57616.52 & 2.79$^{+2.02}_{-0.84}$ & 9.40$^{+5.74}_{-3.92}$ & $1.46 \pm 0.10 \times 10^{43}$ & 1.23 \\
\hline
\multicolumn{6}{c}{2SXPS J020751.5+445038} \\
1 & 55489.38 & 1.09$^{+0.17}_{-0.19}$ & $<1.60$ & $3.97 \pm 0.96 \times 10^{42}$ & 1.07 \\
2 & 55631.12 & 0.98$^{+0.10}_{-0.11}$ & $<1.38$ & $5.27 \pm 0.47 \times 10^{42}$ & 0.69 \\
3 & 55647.52 & 0.95$^{+0.10}_{-0.10}$ & 0.05$^{+0.18}_{-0.05}$ & $5.99 \pm 0.36 \times 10^{42}$ & 0.89 \\
4 & 57333.82 & 0.96$^{+0.07}_{-0.06}$ & $<0.42$ & $5.31 \pm 0.26 \times 10^{42}$ & 1.34 \\
5 & 57334.02 & 0.89$^{+0.14}_{-0.11}$ & $<0.90$ & $3.14 \pm 0.35 \times 10^{42}$ & 0.86 \\
\hline
\multicolumn{6}{c}{2SXPS J215459.6-154104} \\
1 & 54314.06 & 1.50$^{+0.19}_{-0.28}$ & $<0.87$ & $7.77 \pm 0.71 \times 10^{45}$ & 1.06 \\
2 & 54315.0 & 0.58$^{+0.35}_{-0.19}$ & $<3.55$ & $4.26 \pm 0.34 \times 10^{45}$ & 1.27 \\
\hline
    \end{tabular}
    \label{tab:epochbb}
\end{table*}

For two sources, 2SXPS J074553.0+254444 and 2SXPS J020751.5+445038, our fits are generally consistent between epochs, with only small variations in spectral shape implying that the same emission components remain dominant throughout the light curves. This does not necessarily rule out these sources being transients, however, as it is possible that they are in a slow changing state and the more dramatic rise or fall phases have not been observed. For instance, 2SXPS J074553.0+254444 could be consistent with a long lived and slowly fading transient such as a TDE. However, the time period is much longer for 2SXPS J020751.5+445038. It is possible that this source in particular is a recurrent source, such as an AGN with periodic outbursts or a recurrent nova. Unfortunately, the significant time between observations means that from these data, it is impossible to determine exactly what this source might be. 2SXPS J052627.4-211713 and 2SXPS J215459.6-154104 do show more variation between epochs but the individual fits are somewhat poorly constrained. These epochs are therefore consistent within errors and the significance of the apparent variation is unclear.

\subsection{Future observations and analysis}

While it is hard to constrain the behaviour of these transients based on the current data available, it is possible for further observations to better define these sources. Unfortunately, all of these sources were detected at least four years ago and it is likely that any non-recurrent transient behaviour has ceased. Nevertheless, additional X-ray observations could provide key indicators as to the nature of these sources.

\section{Discussion}
\label{sec:discussion}

\subsection{Completeness}

While we have identified 19 transient candidates, it is unclear whether this sample is complete or whether additional transients remain in 2SXPS. To estimate the completeness, we use the GRB catalogue that \textit{Swift} has observed over its mission. 1188 GRBs are included in 2SXPS and are identified by cross-comparison to the \textit{Swift}-XRT GRB catalogue \citep[XRTGRB,][]{Evans09}. Of these, 235 (19.8\%) were included in the InactiveHost and ActiveHost samples, an expected result as a significant number of GRBs are unlikely to have known hosts. In the case of long GRBs, this may be due to them being at high redshifts where their hosts appear extremely dim, while short GRBs may be ejected from their hosts by the high kick velocities induced in their progenitor systems' evolutions \citep[e.g.][]{Mandhai21}. From our classifications in Section \ref{sec:src_class}, 153 GRBs were identified in the combined InactiveHost and ActiveHost samples. This therefore implies a completeness of 65.1\% and that the 2SXPS catalogue may include $\sim30$ previously unidentified transients.

There are several caveats with this estimate, however. GRBs are more luminous and therefore more likely to be picked out as transients. However, other transient populations are more likely to be strongly biased towards inclusion in the InactiveHost and ActiveHost samples and are therefore also more likely to be identified as transients.

There are several possibilities for improving the completeness. Instead of performing the cross-match to putative hosts initially, the flaring analysis detailed in Section \ref{sec:flare_id} could be the first step. This would prevent sources without known hosts being precluded. The host matching step could also be improved by using SIMBAD in addition to NED for a more complete galaxy population. However, as mentioned above, this was not possible for such a large sample size. However, many transients might be lower significance and would therefore be missed. Statistically estimating the effects of Eddington bias more accurately might also help with selecting more effective thresholds. Finally, comparisons could be performed using different catalogues to those used here. In particular, the upcoming release of \textit{eROSITA}'s all sky catalogues would be invaluable as a baseline for such a comparison \citep{erosita}.

\subsection{Future \textit{Swift}-XRT catalogues}

Such improvements could be made with an eye to future catalogues similar to 2SXPS, in particular the live version of the catalogue, LSXPS (Evans et al. in prep), which would allow transient identification in near real time. The live nature of LSXPS offers numerous advantages over a single large catalogue, in addition to identifying transients during their activity. As of May 2022, LSXPS has grown at an average of 48 sources per day compared to 2SXPS. Such a number of sources could be immediately matched to SIMBAD, for instance, improving the purity of the final samples. LSXPS will also likely have the advantage of deeper upper limits in fields that \textit{Swift} has previously observed, as such observations are likely more sensitive than thos performed with \textit{ROSAT} for instance. LSXPS will therefore be more attuned to detecting transients in these fields. It is plausible that a few new extragalactic transients per year could be identified using our methodology with the capability for rapid and effective follow-up.

\section{Conclusions}
\label{sec:conc}

We have searched the 2SXPS catalogue for new transient candidates. By crossmatching to galaxies, we identified the 30599 sources from the original 206335 that were most likely to be extragalactic. Analysis of their light curves and comparisons to other catalogues obtained using \textit{ROSAT} and \textit{XMM-Newton} indicated 706 flaring sources, with 334 appearing to be in inactive galaxies and 372 in active galaxies. We further investigated this final sample and found 19 sources consistent with being newly discovered transient candidates. These putative hosts of these sources are split approximately equally between inactive and active galaxies and the peak X-ray luminosities range from $\sim 10^{42}$ to $10^{47}$ erg s$^{-1}$.

We performed spectral analysis of those sources that had accumulated enough counts and find that eight are best fit with non-thermal models and one with a blackbody, although some sources can be reasonably well fit with multiple models. The properties of these fits are consistent with several classes of X-ray source, such as AGN, GRBs and HMXB. However, our fits using thermal models indicate temperatures that are generally inconsistent with transients such as TDEs.

Several sources had also accumulated enough counts in individual epochs for us to examine how their spectra evolved with time. These sources were found to be consistent within errors in their spectral shape between epochs, but this does not preclude them from indeed being transient sources in long-lived high states or having been consistently observed during a flare. Further observations of these sources are required to determine their exact nature.

The completeness of our methodology was also examined. Based on comparisons to the XRT GRB catalogue, we estimate a completeness of $\sim65$\%. This implies a total of $\sim30$ possible transients in 2SXPS. This bodes well for both the first \textit{eROSITA} data release and the upcoming LSXPS and the possibility of near real time serendipitous discovery of transients.

\section*{Acknowledgements}

This work has made use of: data supplied by the UK \textit{Swift} Science Data Centre at the University of Leicester; data obtained from the 4XMM \textit{XMM-Newton} serendipitous source catalogue compiled by the 10 institutes of the \textit{XMM-Newton} Survey Science Centre selected by ESA; the NASA/IPAC Extragalactic Database (NED), which is funded by the National Aeronautics and Space Administration and operated by the California Institute of Technology; and the SIMBAD database, operated at CDS, Strasbourg, France.

RAJEF acknowledges support from the UKSA, the STFC and the European Union’s Horizon 2020 Programme under the AHEAD2020 project (grant agreement number 871158). PAE acknowledges UKSA support.

The authors thank Mat Page and Mike Goad for their valuable comments and suggestions and the anonymous referee for their very appreciated feedback.

\section*{Data Availability}

The 2SXPS catalogue and associated data are available from the UKSSDC at \url{https://www.swift.ac.uk/2SXPS/}. Other data used in this paper are available from the urls in the relevant footnotes.



\bibliographystyle{mnras}
\bibliography{bibs/2sxps,bibs/software,bibs/swift,bibs/TDEs} 




\appendix

\newpage
\section{Source light curves}
\label{sec:lc}

In this Appendix, we present the 2SXPS light curves of our transient candidates.

\begin{figure*}
    \centering
    \includegraphics[width=5cm]{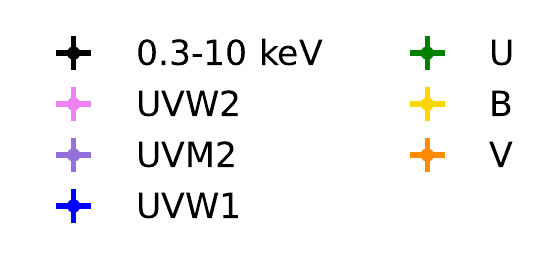}
    \includegraphics[width=\columnwidth]{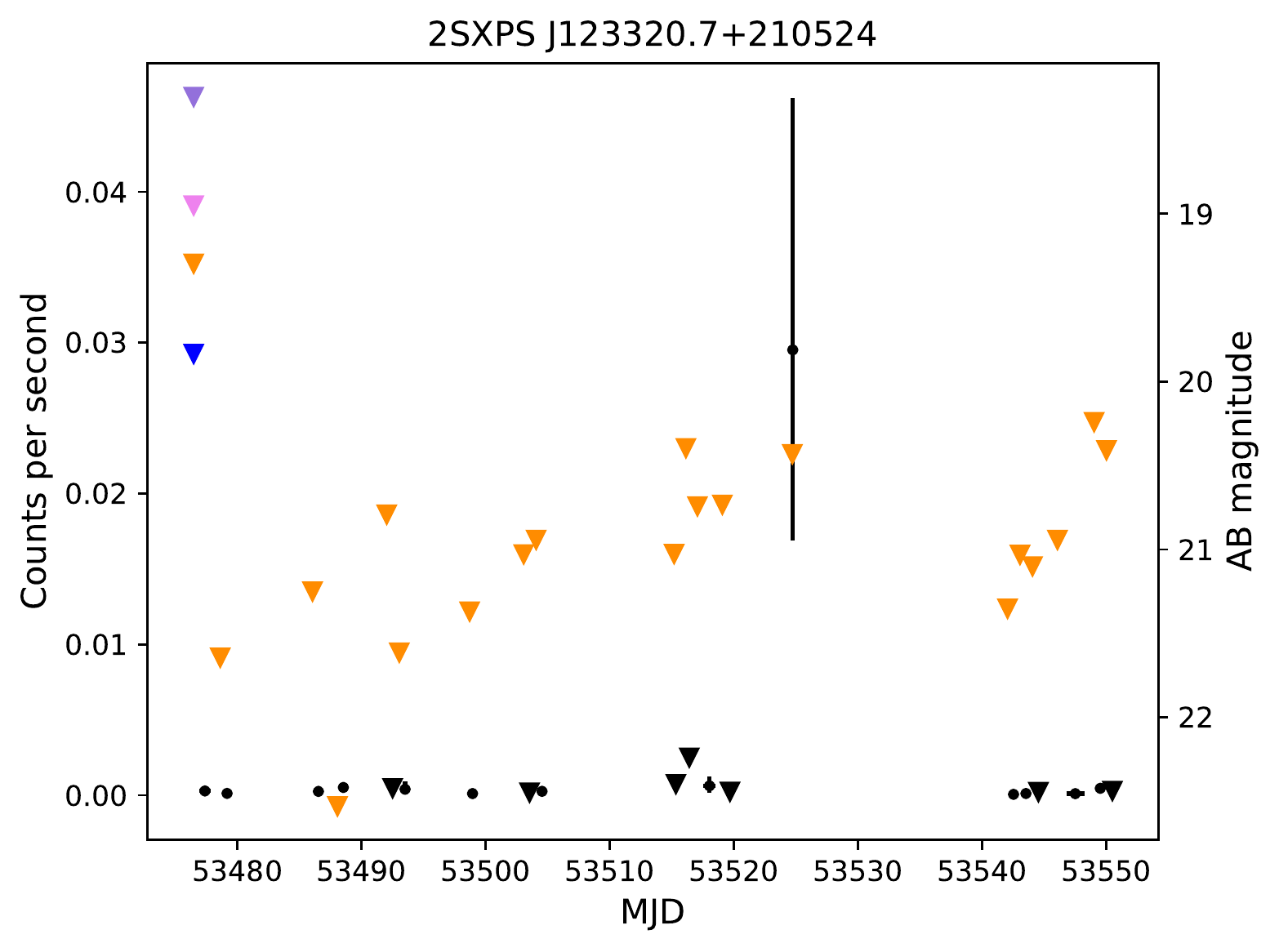}
    \includegraphics[width=\columnwidth]{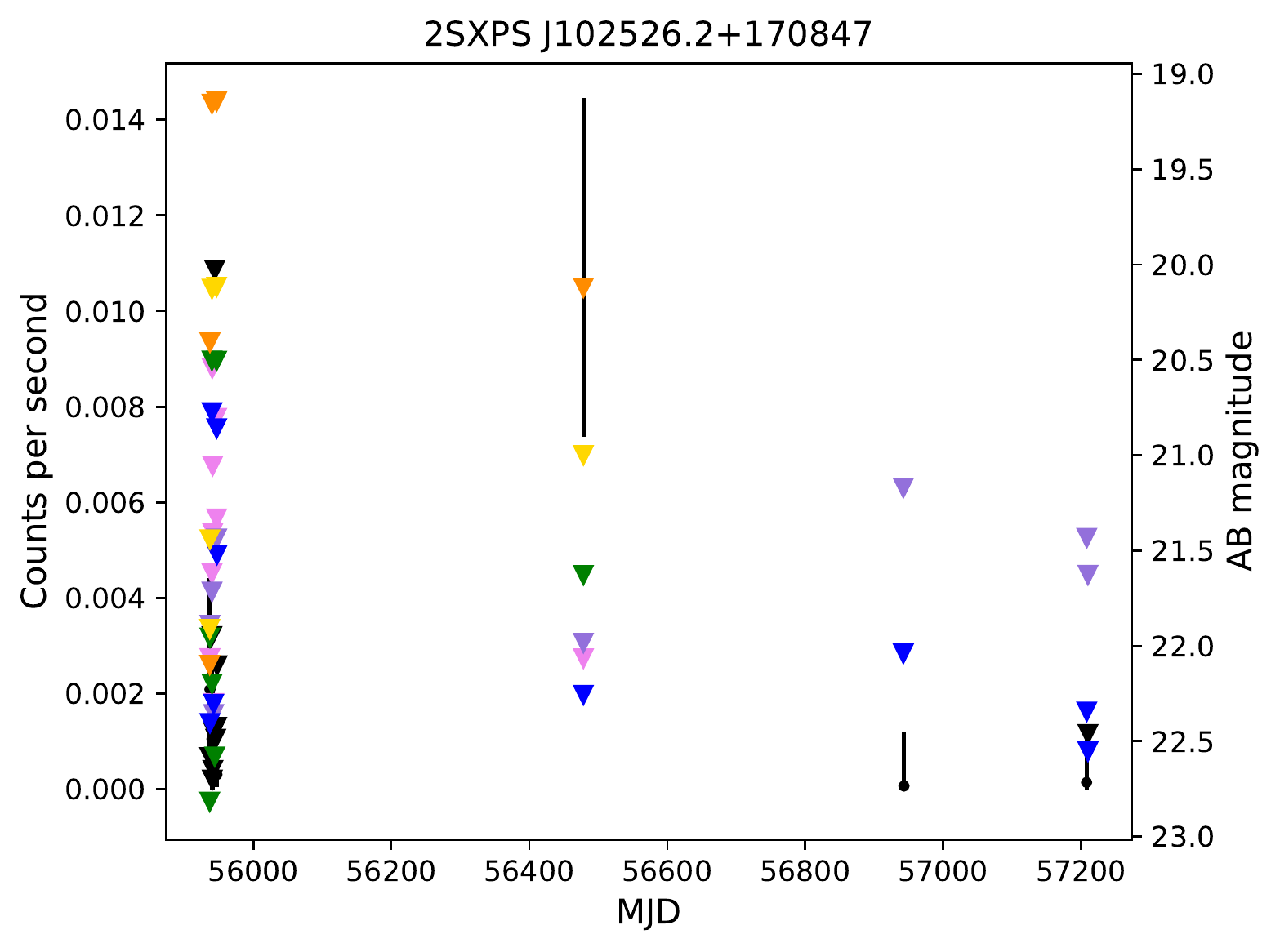}
    \includegraphics[width=\columnwidth]{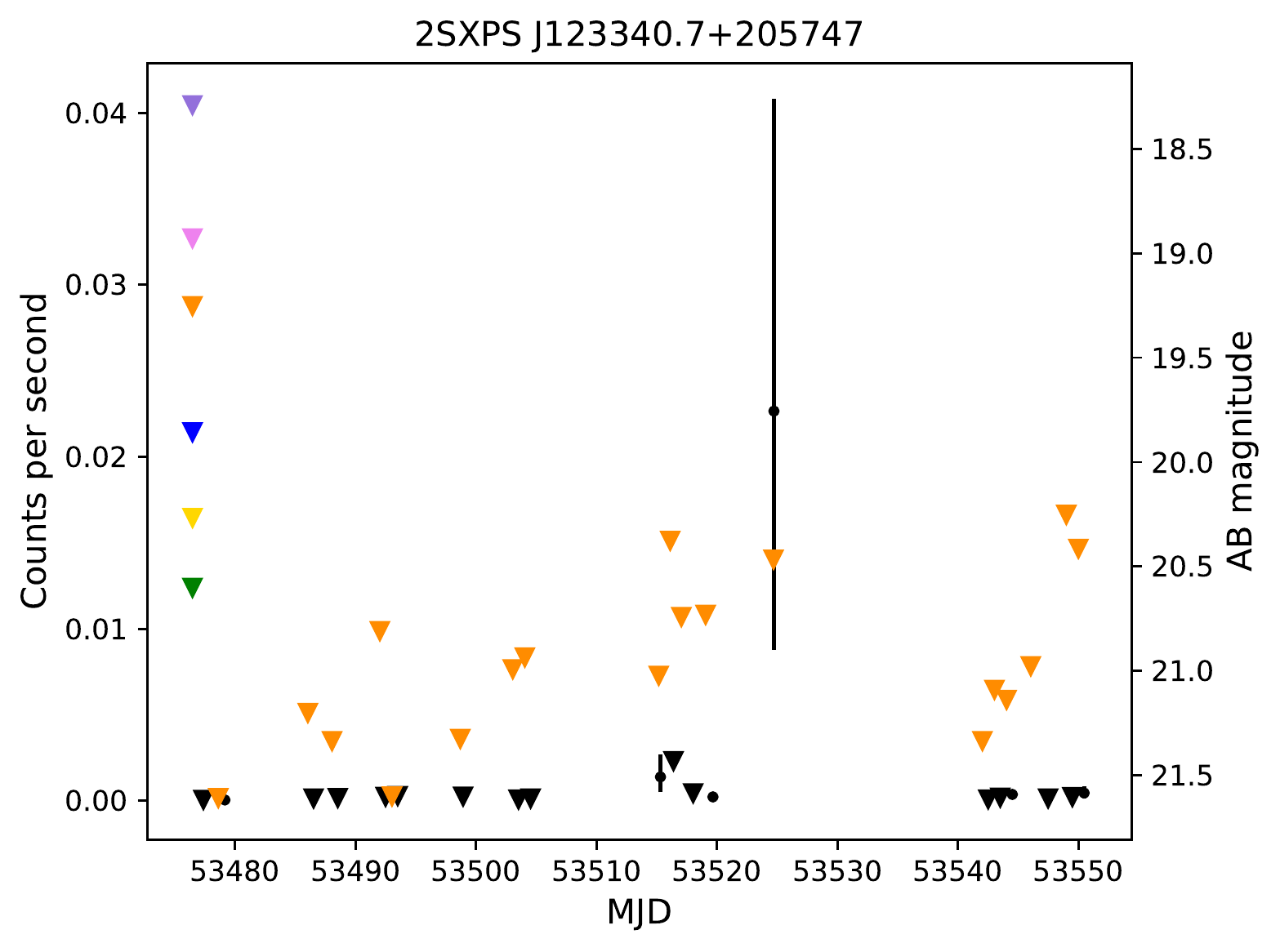}
    \includegraphics[width=\columnwidth]{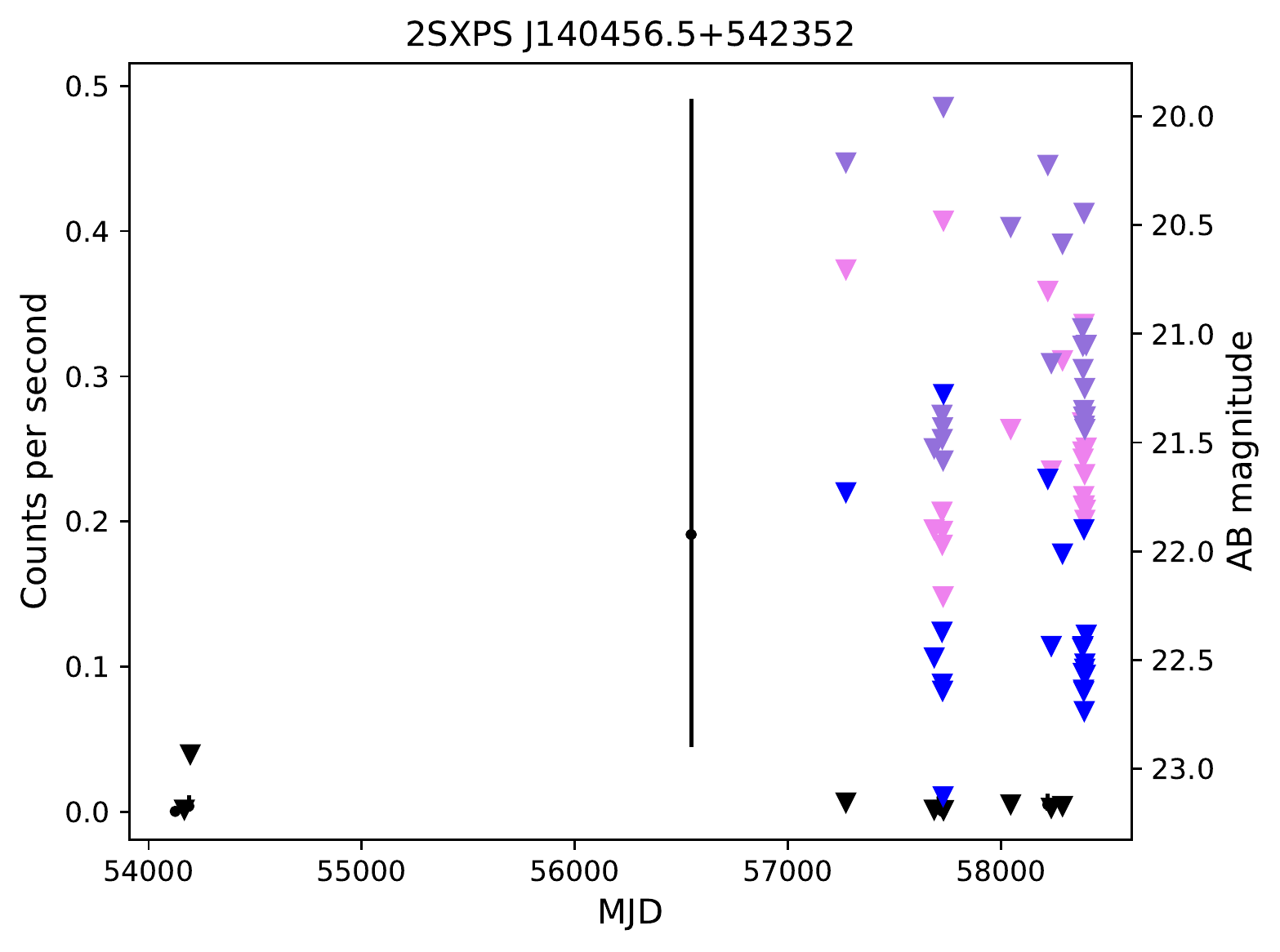}
    \includegraphics[width=\columnwidth]{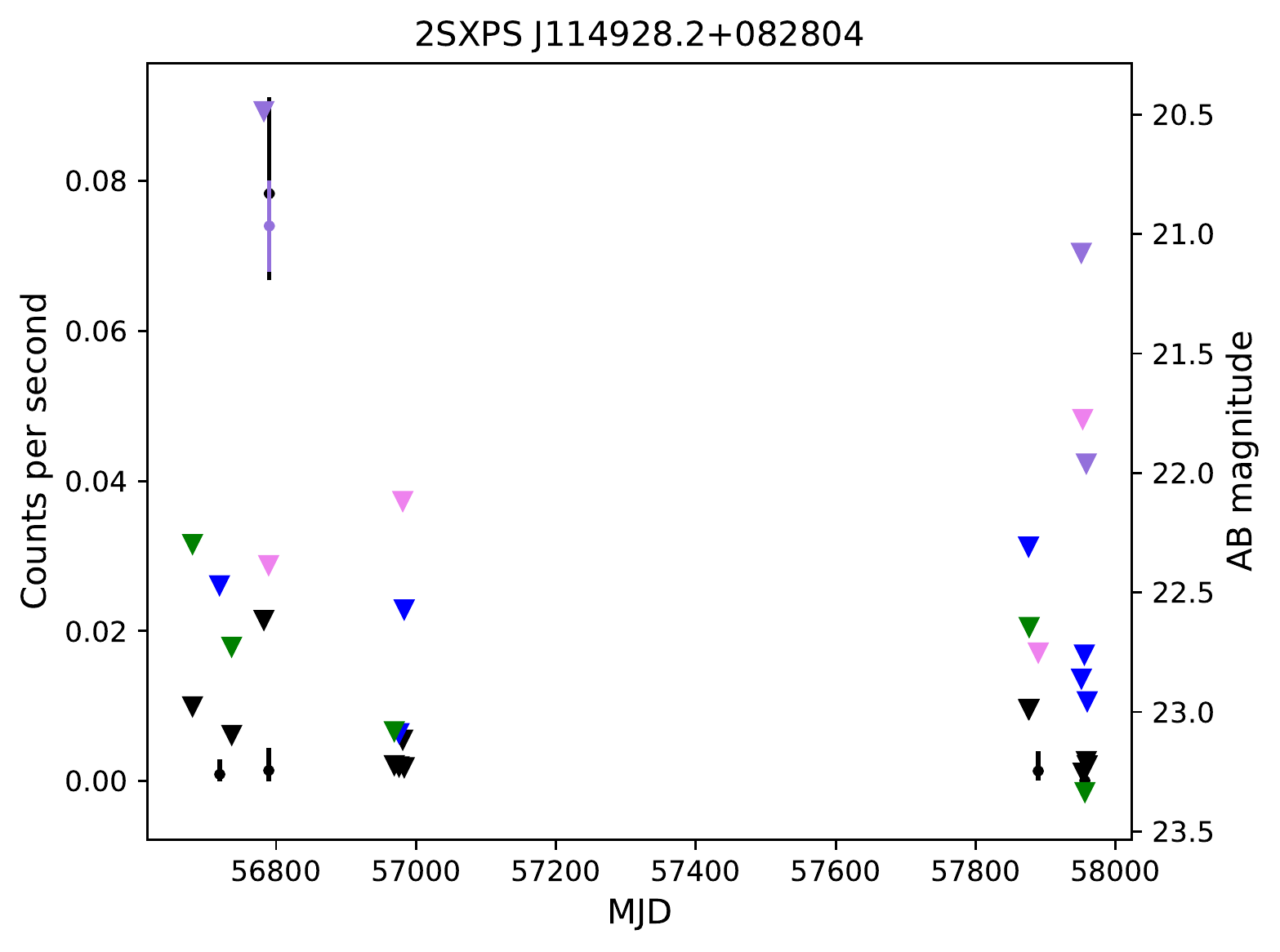}
    \caption{{\it Swift} XRT and UVOT light curves of each of the faint transient candidates. The legend applies to all panels. Note the varying scales on each axis.}
    \label{fig:faintlc}
\end{figure*}

\begin{figure*}
    \centering
    \includegraphics[width=\columnwidth]{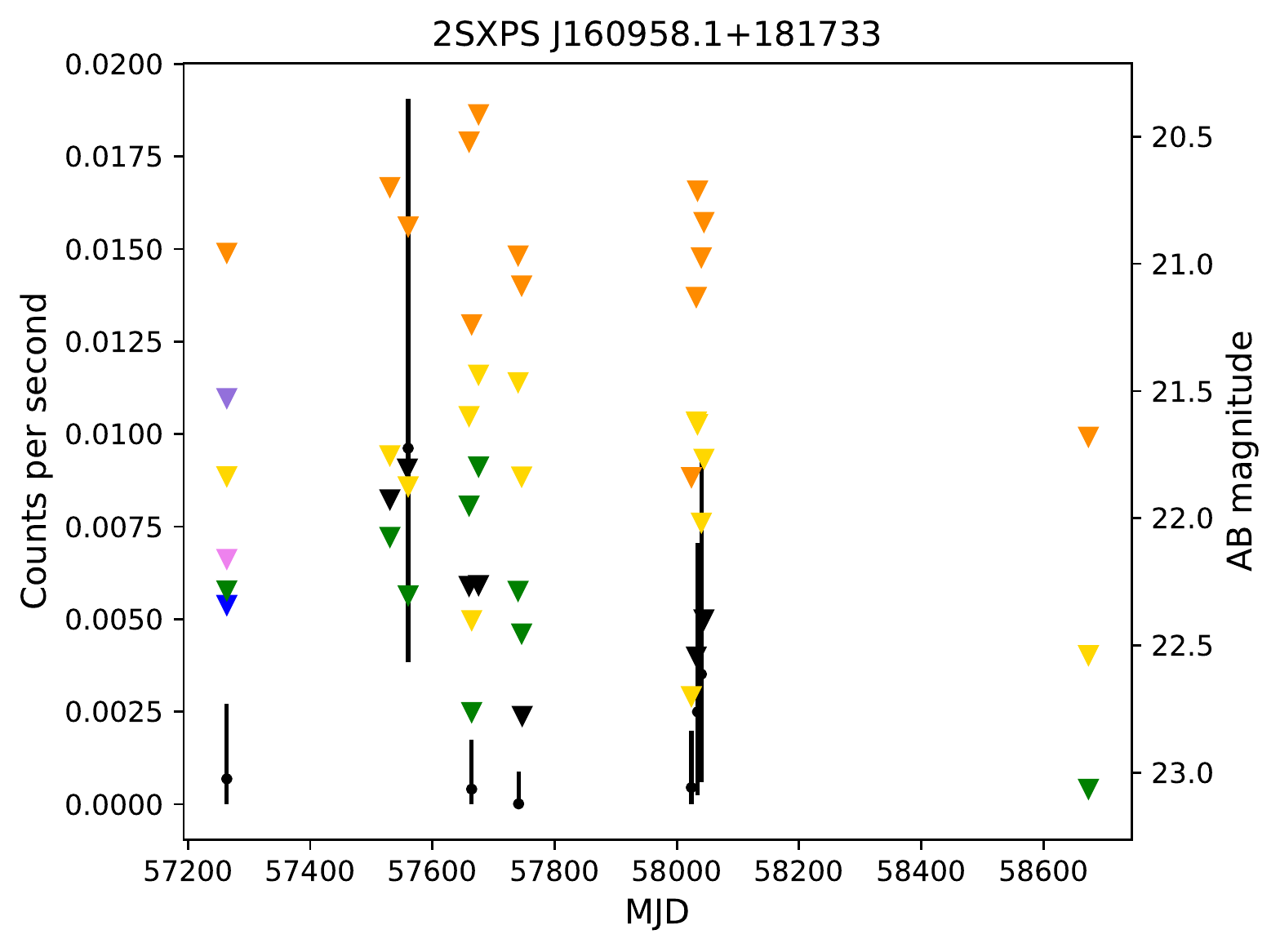}
    \includegraphics[width=\columnwidth]{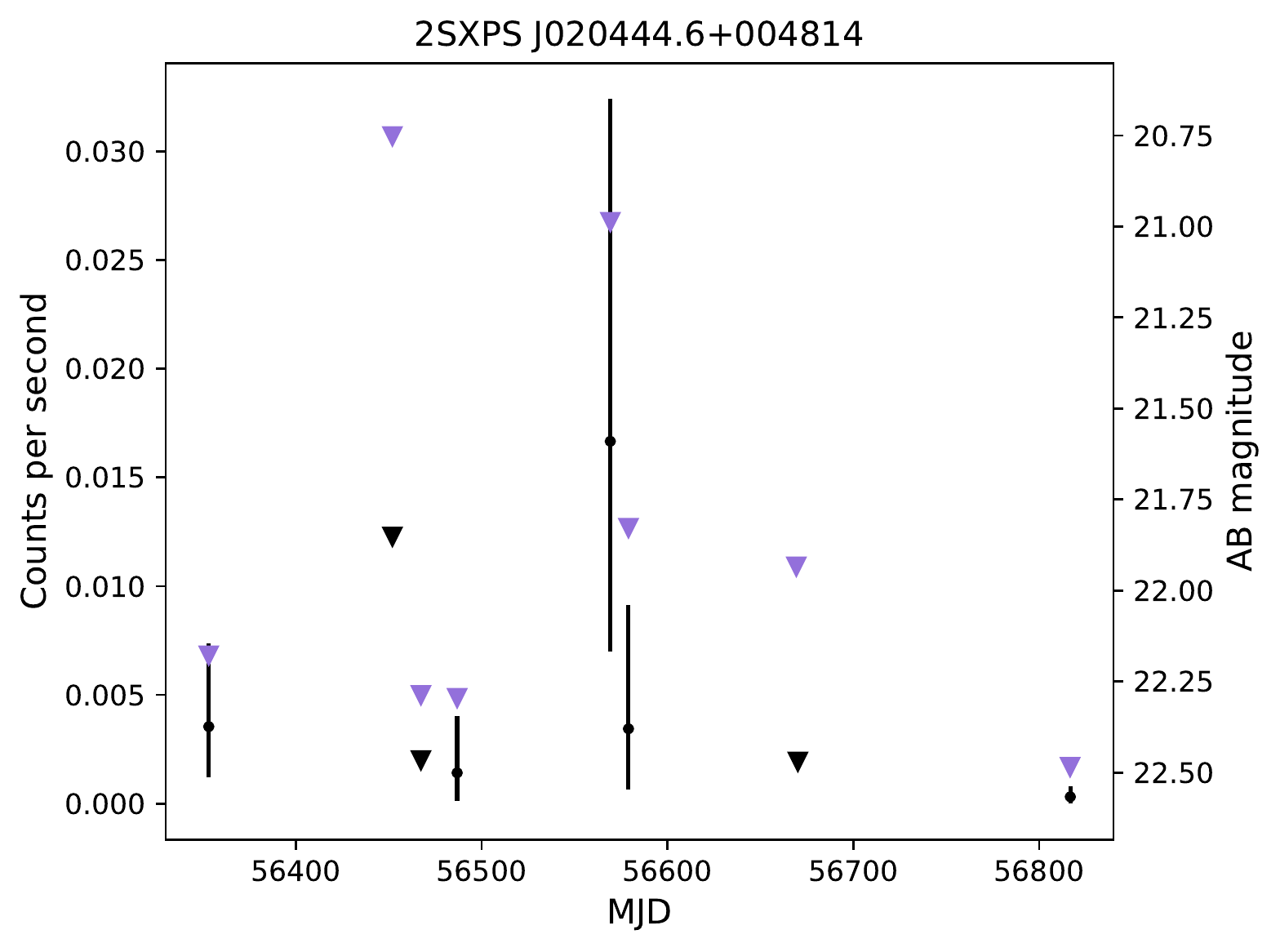}
    \includegraphics[width=\columnwidth]{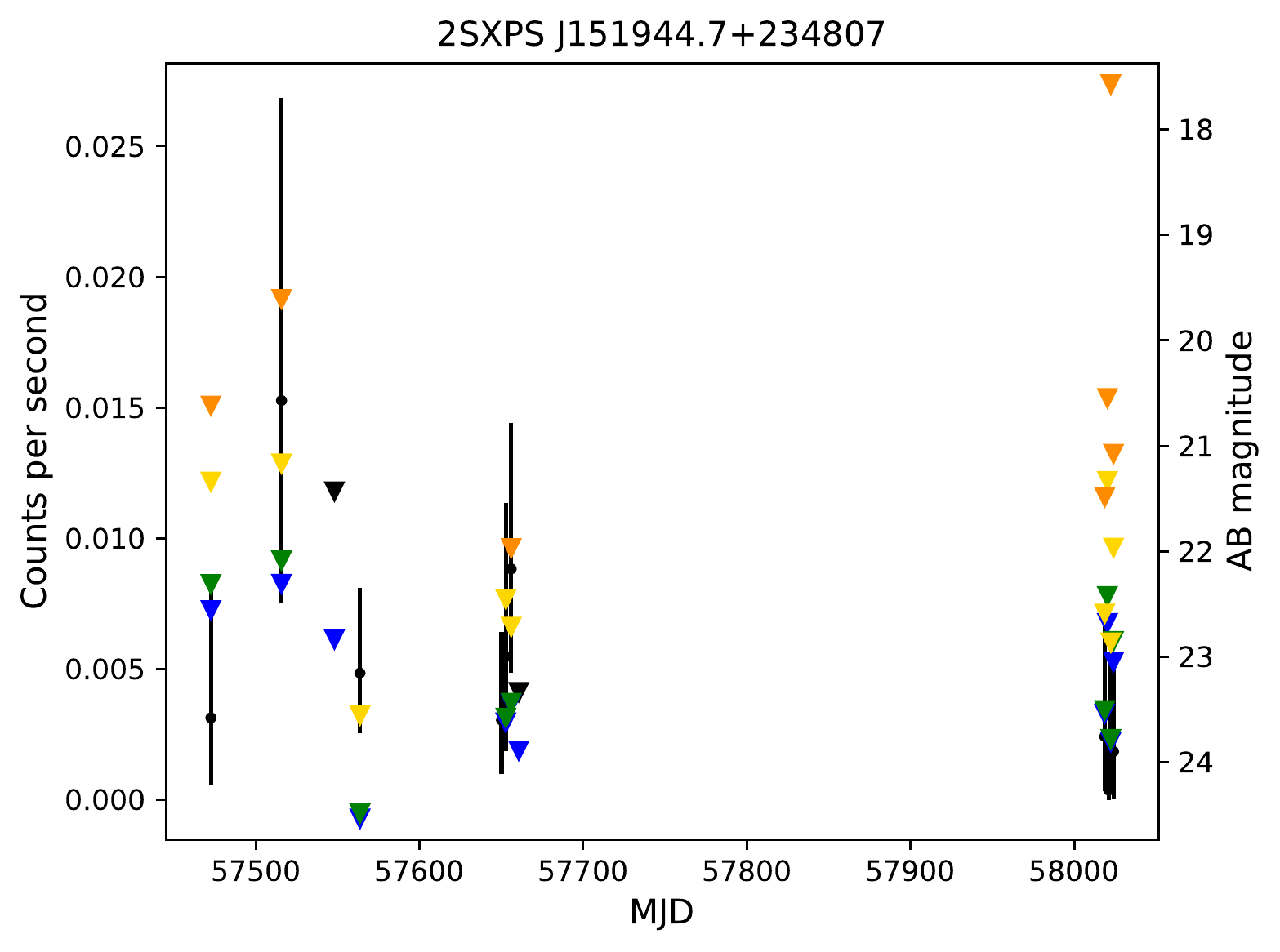}
    \includegraphics[width=\columnwidth]{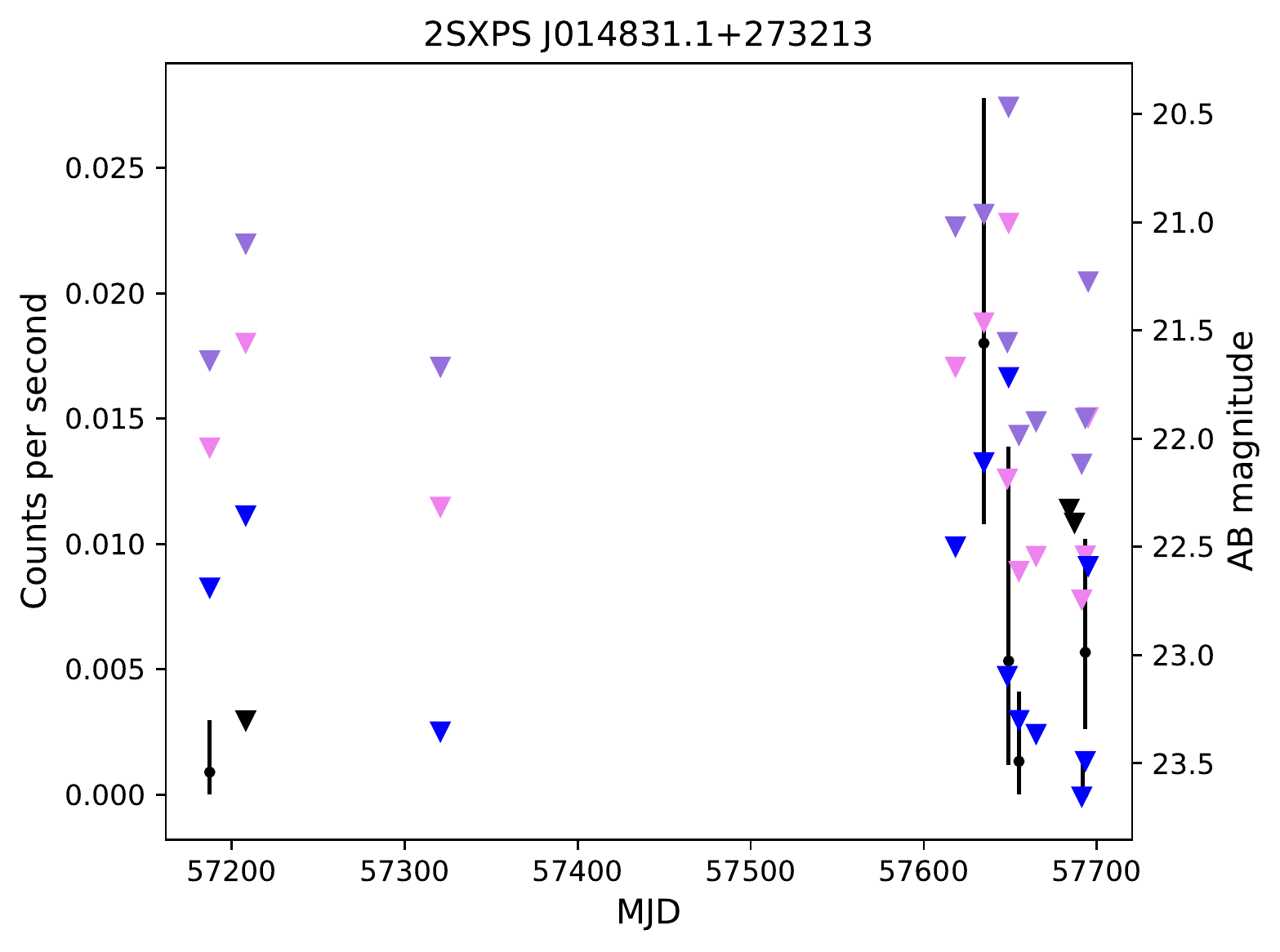}
    \includegraphics[width=\columnwidth]{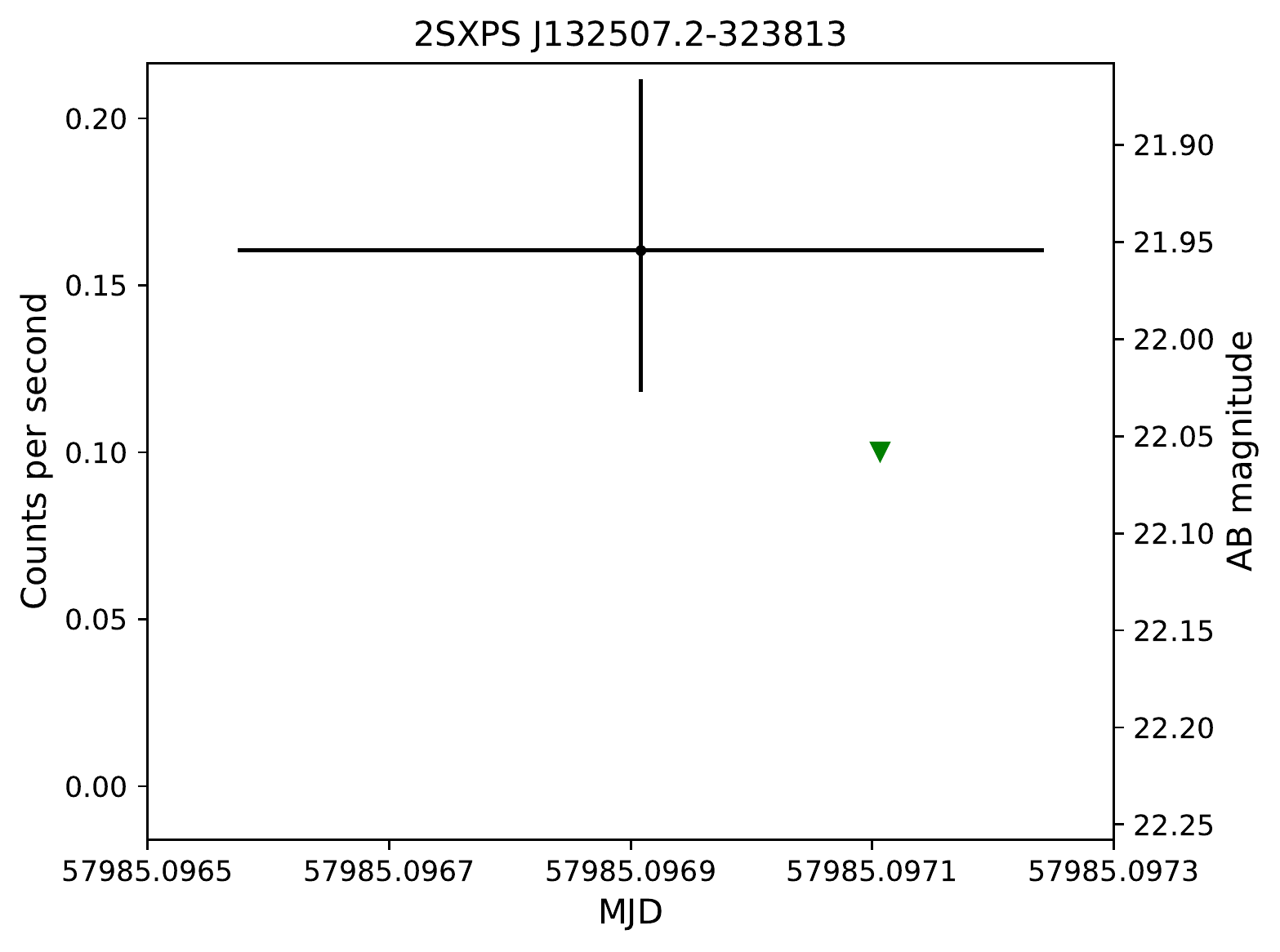}
    \contcaption{}
\end{figure*}

\begin{figure*}
    \centering
    \includegraphics[width=5cm]{figs/legend.pdf}
    \includegraphics[width=\columnwidth]{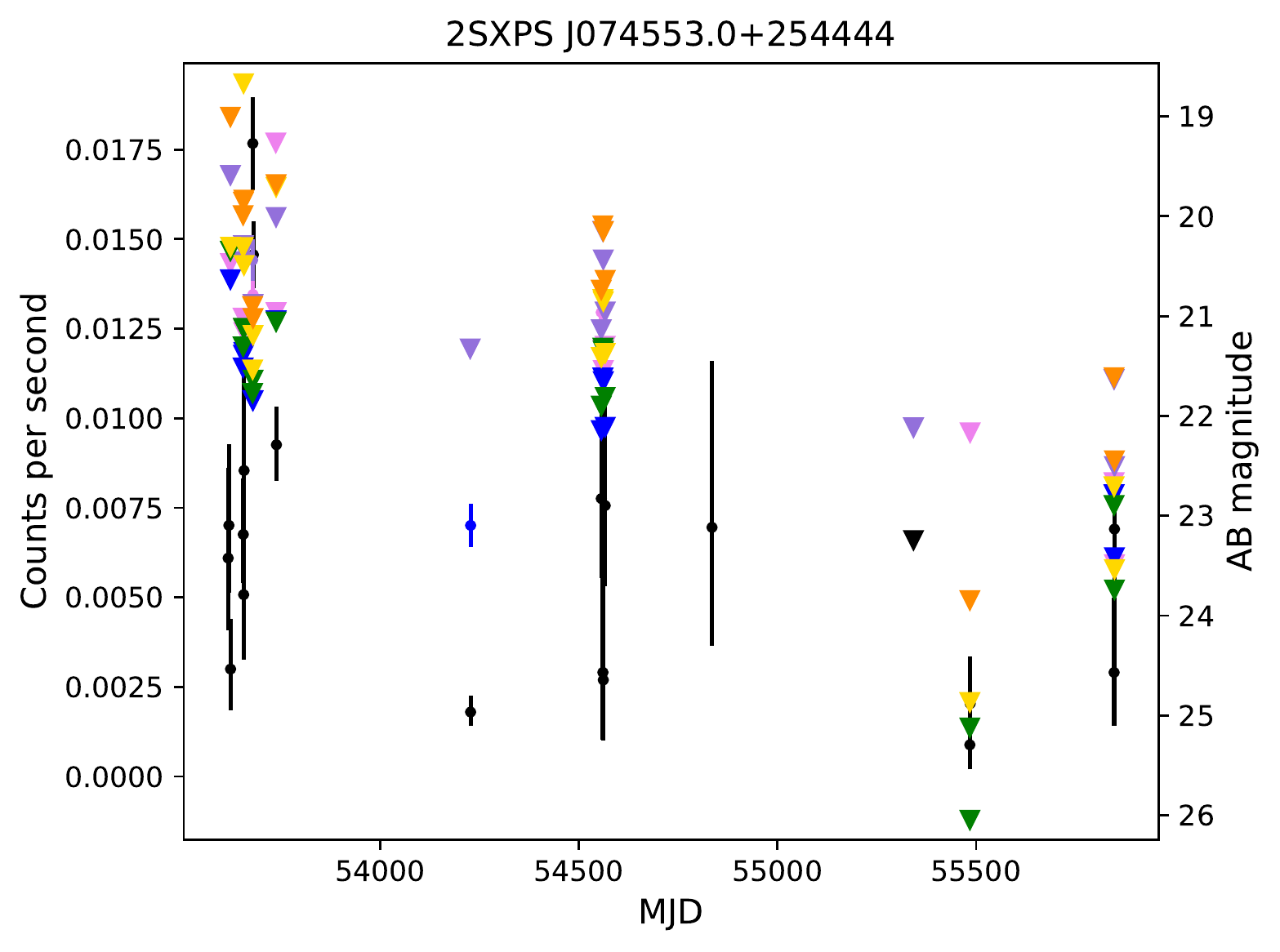}
    \includegraphics[width=\columnwidth]{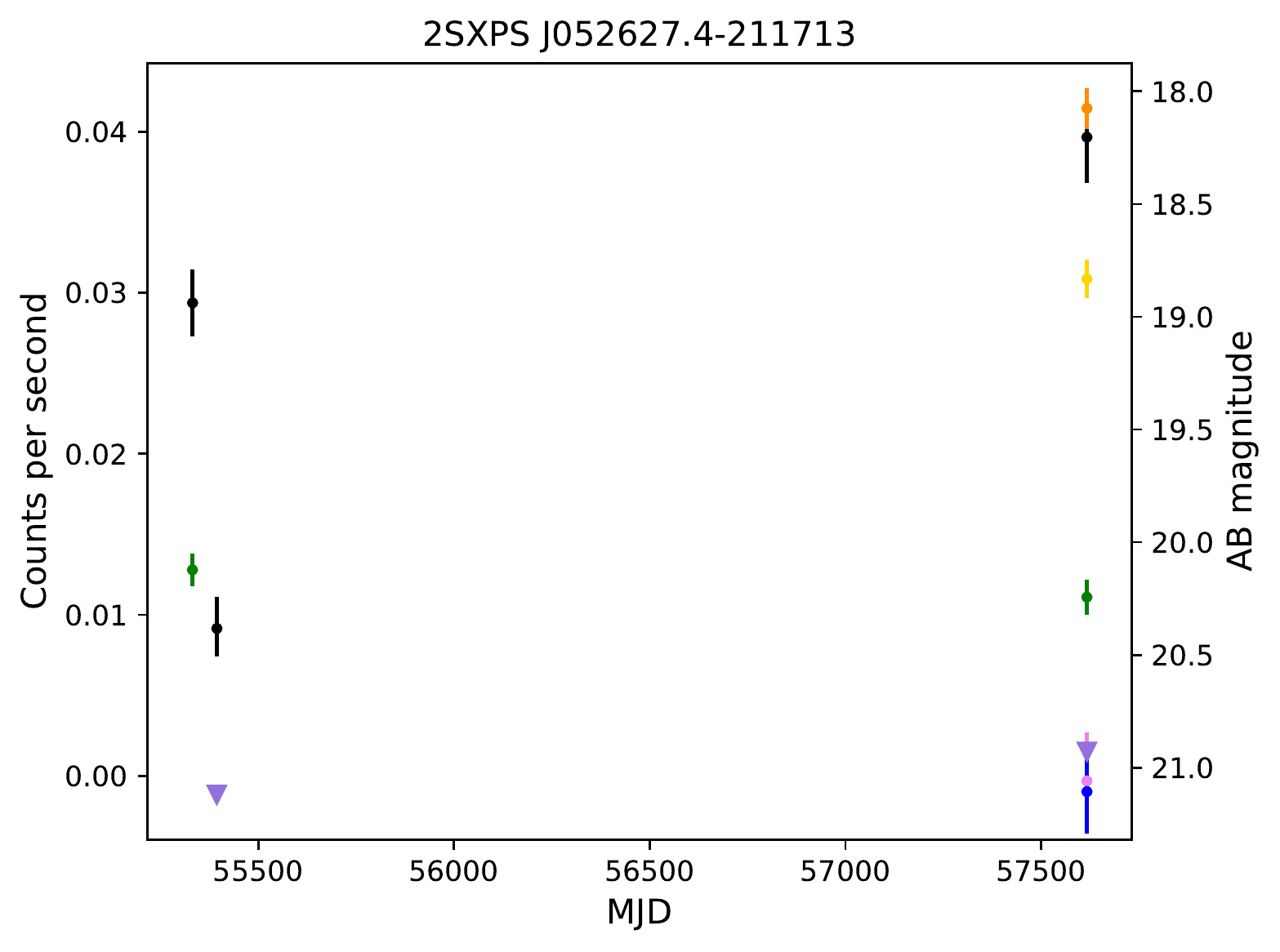}
    \includegraphics[width=\columnwidth]{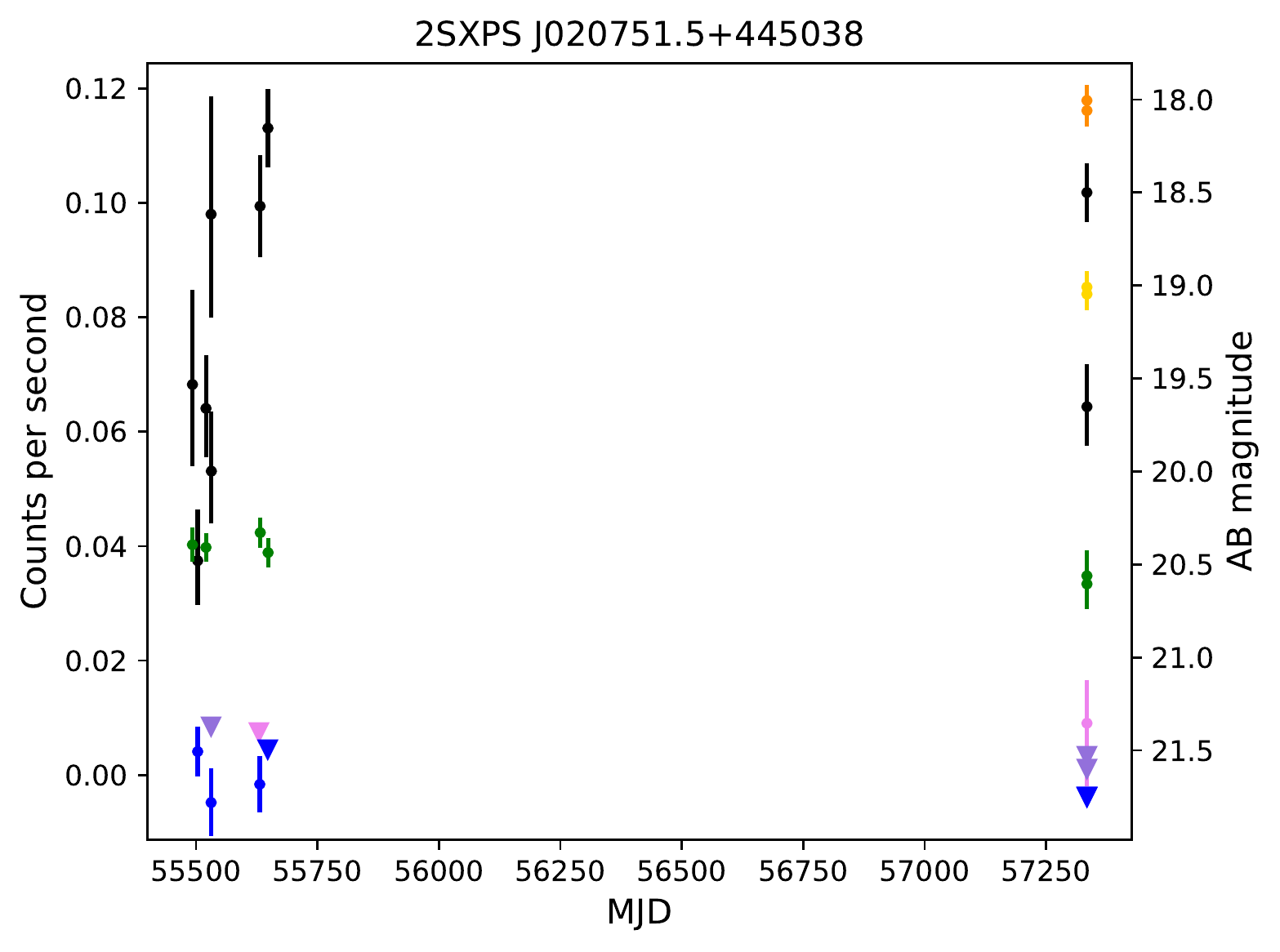}
    \includegraphics[width=\columnwidth]{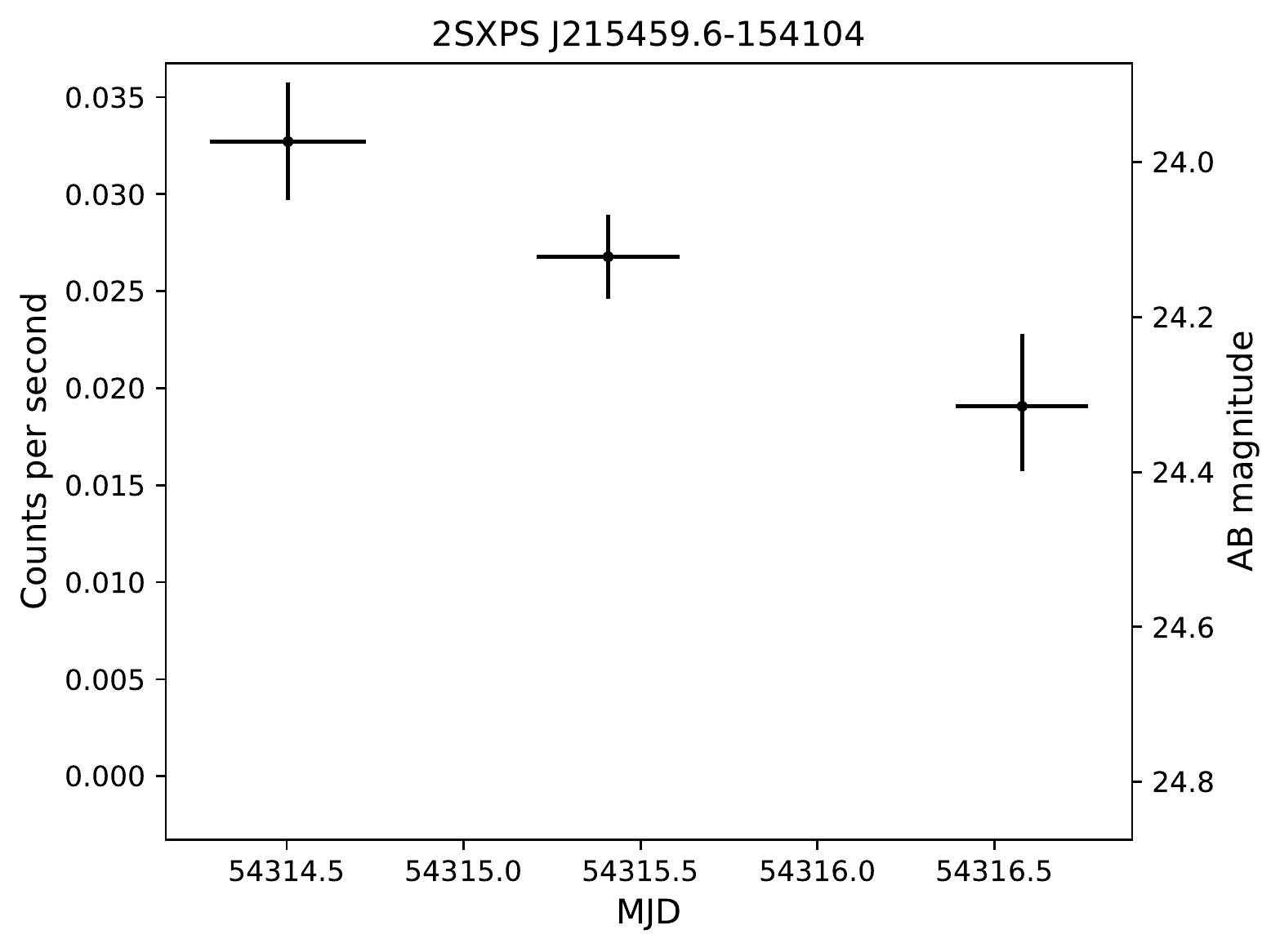}
    \includegraphics[width=\columnwidth]{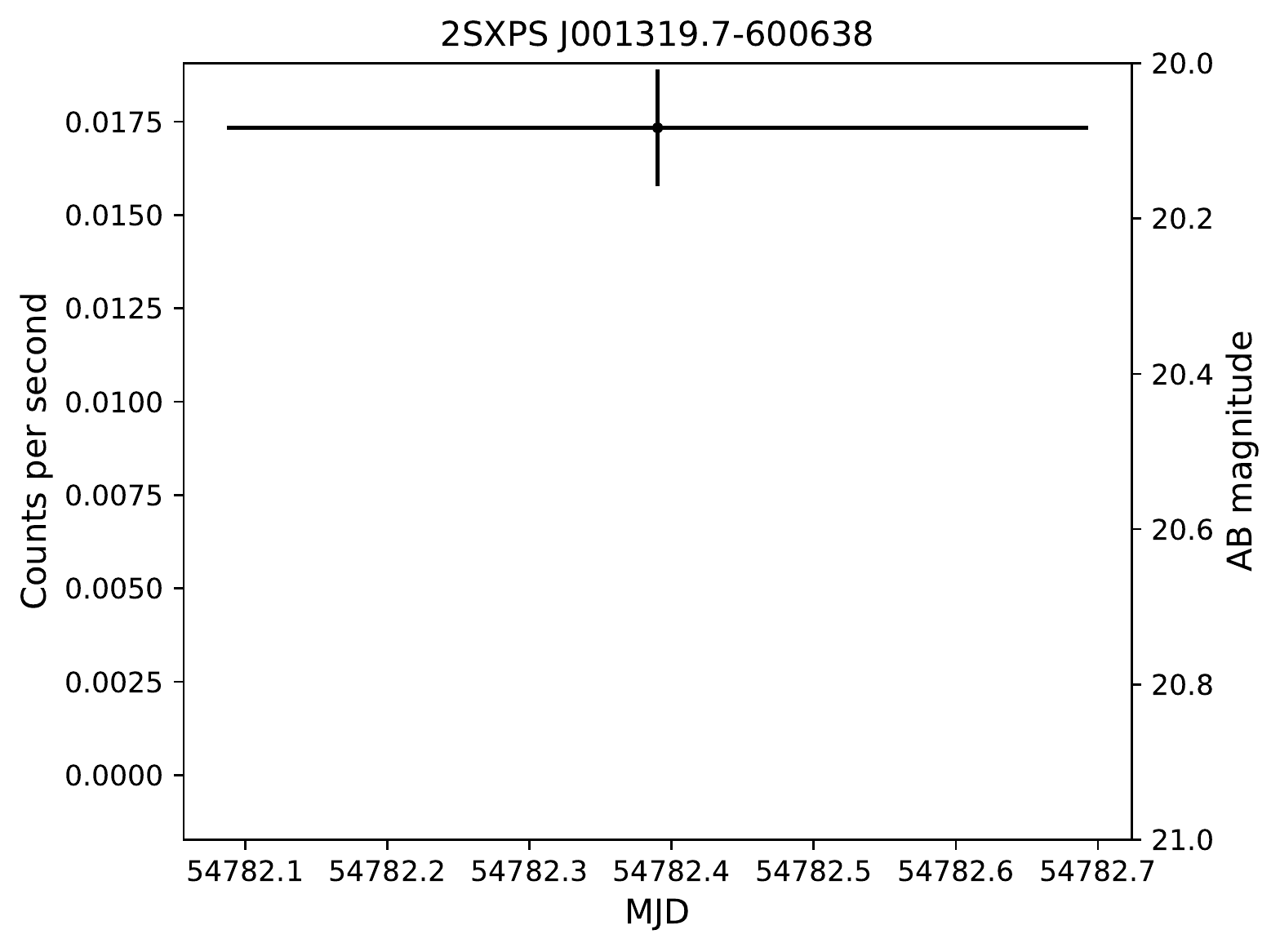}
    \caption{The XRT and UVOT light curves of each of the bright transient candidates. The legend applies to all panels. Note the varying scales on each axis.}
    \label{fig:brightlc}
\end{figure*}

\begin{figure*}
    \centering
    \includegraphics[width=\columnwidth]{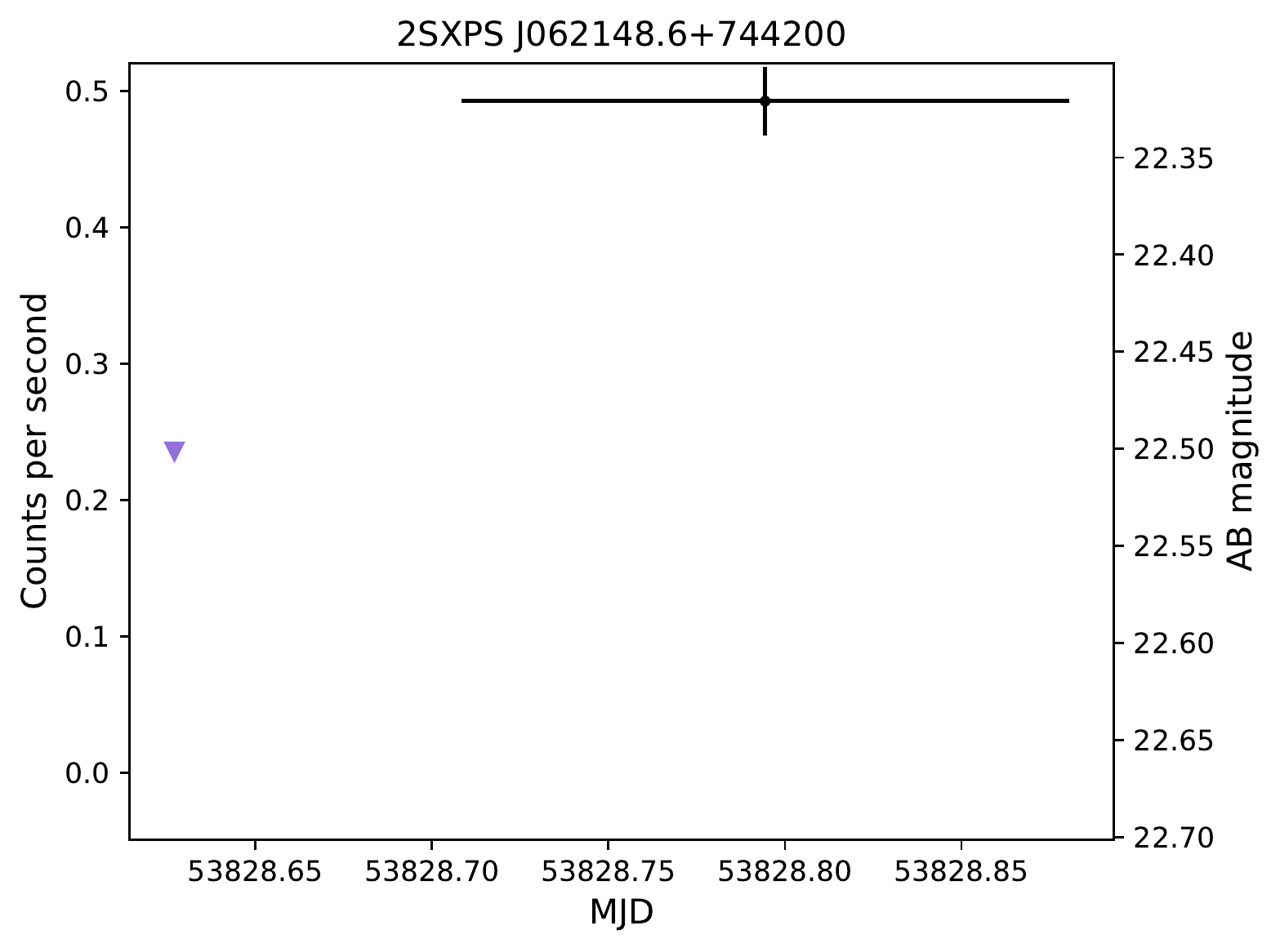}
    \includegraphics[width=\columnwidth]{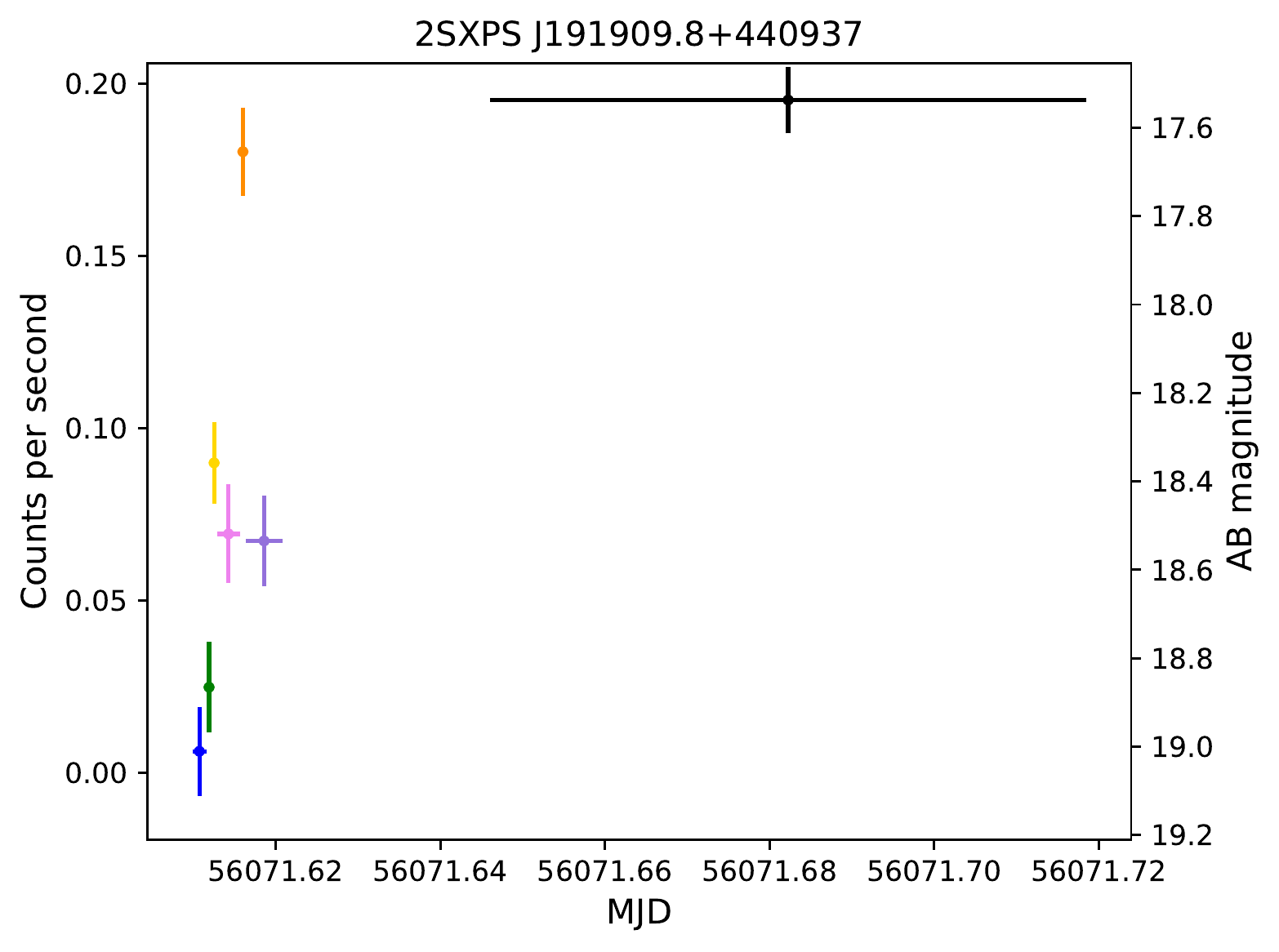}
    \includegraphics[width=\columnwidth]{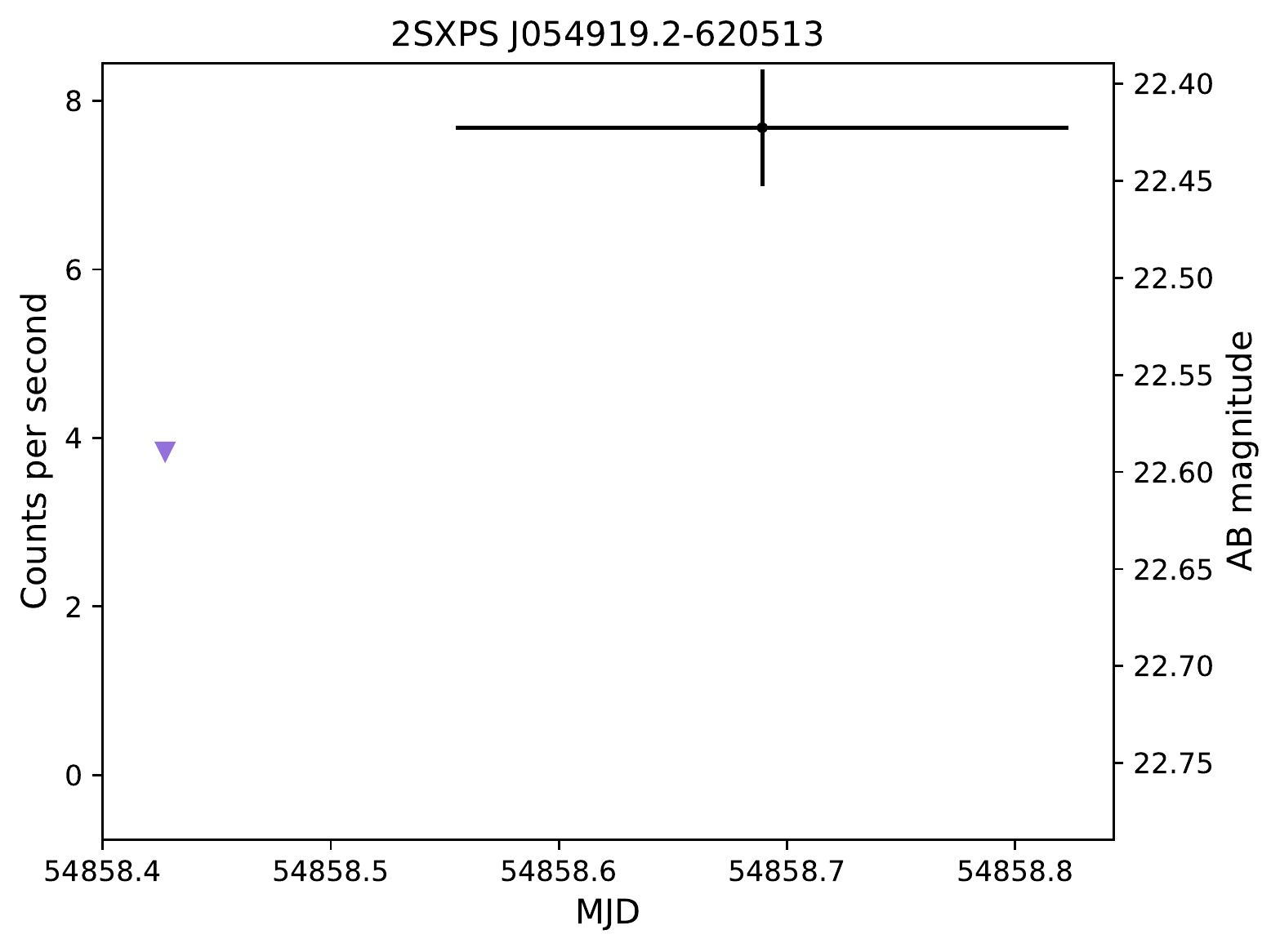}
    \includegraphics[width=\columnwidth]{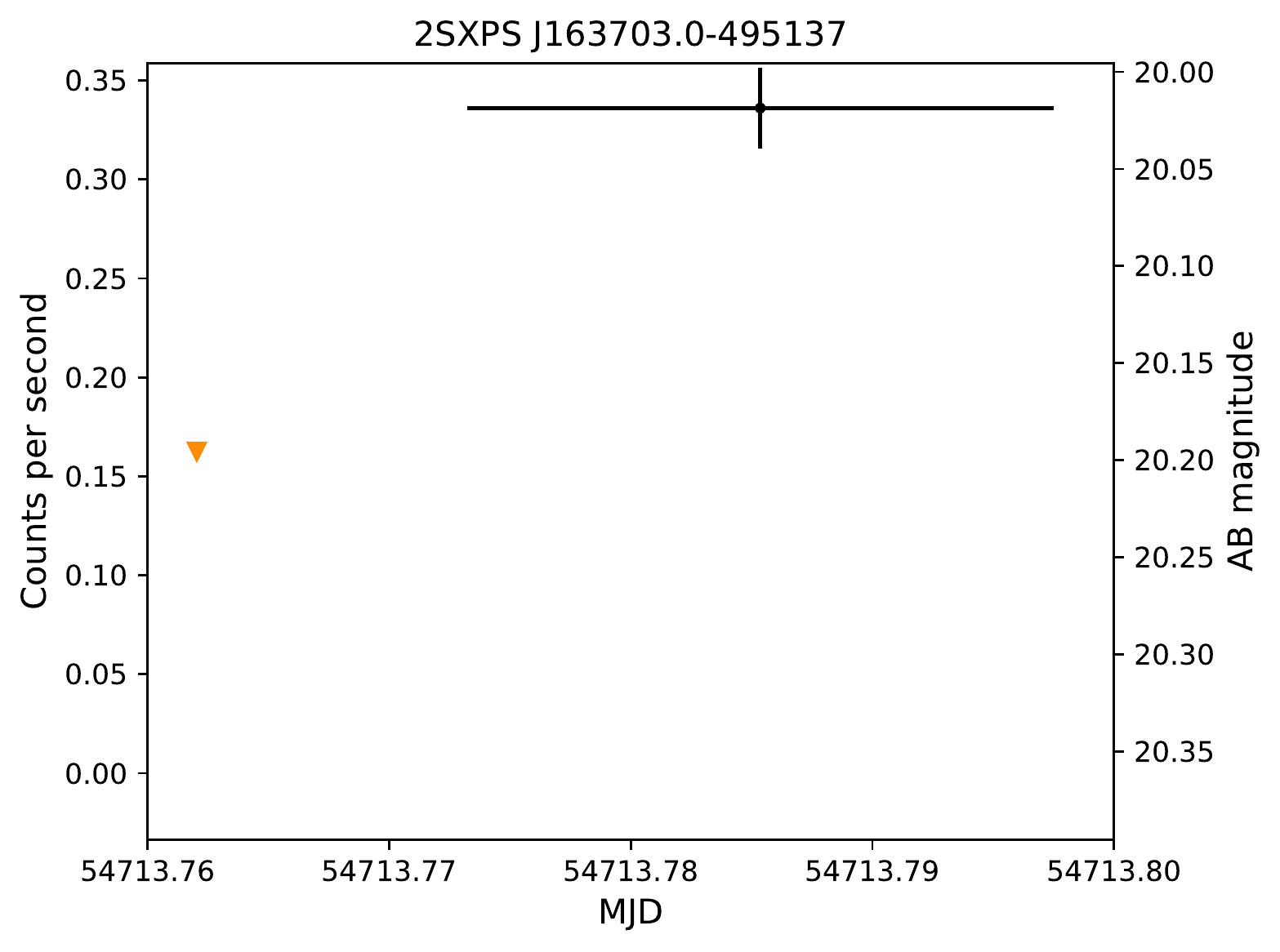}
    \contcaption{}
\end{figure*}



\bsp	
\label{lastpage}
\end{document}